\documentclass[twocolumn,twocolappendix]{aastex631}
\usepackage{amsmath}

\received{September 8, 2021}
\revised{September 19, 2025}
\accepted{September 21, 2025}

\submitjournal{ApJL}

\graphicspath{{./}{figures/}}

\begin{document}

\title{Multi-spacecraft Measurements of the Evolving Geometry of the Solar Alfv\'en Surface Over Half a Solar Cycle}

\correspondingauthor{Samuel T. Badman}
\email{samuel.badman@cfa.harvard.edu}

\author[0000-0002-6145-436X]{Samuel T. Badman}
\affiliation{Center for Astrophysics $\vert$ Harvard \& Smithsonian \\
Cambridge, MA 02138, USA}

\author[0000-0002-7728-0085]{Michael L. Stevens}
\affiliation{Center for Astrophysics $\vert$ Harvard \& Smithsonian \\
Cambridge, MA 02138, USA}

\author[0000-0002-1989-3596]{Stuart D. Bale}
\affil{Physics Department, University of California, Berkeley, CA 94720-7300, USA}
\affil{Space Sciences Laboratory, University of California, Berkeley, CA 94720-7450, USA}

\author[0000-0002-8748-2123]{Yeimy J. Rivera}
\affiliation{Center for Astrophysics $\vert$ Harvard \& Smithsonian \\
Cambridge, MA 02138, USA}

\author[0000-0001-6038-1923]{Kristopher G. Klein}
\affiliation{Lunar and Planetary Laboratory, University of Arizona, Tucson, AZ 85721, USA}

\author[0000-0001-6692-9187]{Tatiana Niembro}
\affiliation{Center for Astrophysics $\vert$ Harvard \& Smithsonian \\
Cambridge, MA 02138, USA}

\author[0000-0002-7174-6948]{Rohit Chhiber}
\affiliation{Department of Physics and Astronomy, University of Delaware
Newark, DE 19716, USA}
\affiliation{Heliophysics Science Division, NASA Goddard Space Flight Center, Greenbelt, Maryland, 20771, USA}

\author{Ali Rahmati} 
\affil{Space Sciences Laboratory, University of California, Berkeley, CA 94720-7450, USA}

\author[0000-0002-7287-5098]{Phyllis L. Whittlesey} 
\affil{Space Sciences Laboratory, University of California, Berkeley, CA 94720-7450, USA}

\author[0000-0002-0396-0547]{Roberto Livi} 
\affil{Space Sciences Laboratory, University of California, Berkeley, CA 94720-7450, USA}

\author[0000-0001-5030-6030]{Davin E. Larson} 
\affil{Space Sciences Laboratory, University of California, Berkeley, CA 94720-7450, USA}

\author[0000-0002-5982-4667]{Christopher J. Owen}
\affiliation{Mullard Space Science Laboratory, University College London, Holmbury St. Mary, Dorking, Surrey, RH5 6NT, UK}

\author[0000-0002-5699-090X]{Kristoff W. Paulson}
\affiliation{Center for Astrophysics $\vert$ Harvard \& Smithsonian \\
Cambridge, MA 02138, USA}

\author[0000-0002-7572-4690]{Timothy S. Horbury}
\affiliation{Department of Physics, Imperial College London, London, SW7 2BW, UK}

\author{Jean Morris}
\affiliation{Department of Physics, Imperial College London, London, SW7 2BW, UK}

\author{Helen O'Brien}
\affiliation{Department of Physics, Imperial College London, London, SW7 2BW, UK}

\author[0000-0002-1628-0276]{Jean-Baptiste Dakeyo}
\affil{Space Sciences Laboratory, University of California, Berkeley, CA 94720-7450, USA}

\author[0000-0003-1138-652X]{Jaye L. Verniero}
\affiliation{Heliophysics Science Division, NASA Goddard Space Flight Center, Greenbelt, Maryland, 20771, USA}

\author[0000-0002-7365-0472]{Mihailo Martinovic}
\affiliation{Lunar and Planetary Laboratory, University of Arizona, Tucson, AZ 85721, USA}

\author[0000-0002-1573-7457]{Marc Pulupa}
\affiliation{Space Sciences Laboratory, University of California, Berkeley, CA 94720-7450, USA}

\author[0000-0002-5456-4771]{Federico Fraschetti}
\affiliation{Center for Astrophysics $\vert$ Harvard \& Smithsonian \\
Cambridge, MA 02138, USA}

\begin{abstract}
The geometry of a star's Alfv\'en surface determines stellar angular momentum loss, separates a causally distinct ``corona'' and stellar wind, and potentially affects exoplanetary habitability. The solar Alfv\'en surface is the only such structure that is directly measurable and since 2021, has been routinely measured \textit{in situ} by NASA's Parker Solar Probe (Parker). We use these unique measurements in concert with Solar Orbiter and L1 in situ data spanning the first half of the Solar\,Cycle\,25 in time and from 0.045\,--\,1\,au in heliocentric distance to develop a radial scaling technique to estimate the morphology of the Alfv\'en surface from measurements of the solar wind speed and local Alfv\'en speed. We show that accounting for solar wind acceleration and mass flux is necessary to achieve reasonable agreement between the scaled location of the Alfv\'en surface and the locations of direct crossings measured by Parker. We produce continuous 2D equatorial cuts of the Alfv\'en surface over half a Solar\,Cycle (ascending phase and maximum). Parker’s earliest crossings clipped outward extrusions, many of which are likely transient related, while more recently Parker has unambiguously sampled deep sub-Alfv\'enic flows. We analyze the average altitude, departure from spherical symmetry, and surface roughness, finding that all are positively correlated to solar activity. For the current modest Solar\,Cycle, the height varies up to 30\% which corresponds to a near-doubling in angular momentum loss per unit mass loss.  
\end{abstract}


\section{Introduction}\label{sec:intro}



As the solar wind accelerates from an initially static state to many hundreds of km\,s$^{-1}$ far from the Sun, it must pass through several critical speed transitions. In hydrodynamics, this is simply the sonic point \citep{Parker1958, Parker1960}, where the sound speed equals the flow speed.
When modeled as a magnetohydrodynamic fluid, there are critical points associated with the fast and slow magnetosonic speeds, as well as the Alfv\'en speed \citep{Weber1967}. 

Due to the enormous importance of Alfv\'en waves and more broadly of Alfv\'enic fluctuations and turbulence in dynamics and evolution of the corona \citep[e.g.][]{Hollweg1978,Velli1993,Tomczyk2007} and near-Sun solar wind \citep[e.g. ][]{BelcherDavis1971,Bale2019,Rivera2024a}, the Alfv\'en critical point has received particular interest. In recent years, with detailed knowledge of the near-Sun solar wind topology, the concept of a single critical point in a spherically symmetric system \citep{Weber1967} has been relaxed in favor of discussing a non-uniform ``Alfv\'en surface'' \citep{Kasper2021} or even a 3D ``Alfv\'en region'' \citep{Chhiber2022}.

No matter its dimensionality, physical interest in this boundary location stems both from its nature as a scale height for the radial evolution of different solar wind streams, and from open questions surrounding different operative physical processes above and below it. As a scale height, it specifies the rate of angular momentum flux loss per unit mass along a given streamline \citep[e.g.][]{Weber1967, Finley2019, Dakeyo2024a} and is a key inflection point in models and empirical observations of Alfv\'en wave energy flux \citep{Cranmer2023,Ruffolo2024}. It is also an interesting boundary intrinsically in that it separates a causally connected sub-Alfv\'enic volume in which information can propagate from any point to another \citep[including inwards in the solar rest frame;][]{Tenerani2016}, from a super-Alfv\'enic wind where information cannot be transmitted inwards, but is always advected outwards. This has led to the Alfv\'en surface being connected to physical transitions such as the height of helmet streamers  \citep{Zhao2010} and closed loops \citep[which has even been connected to exoplanet habitability;][]{Atkinson2024}, regions of preferential minor ion heating \citep{Kasper2019}, and to turbulent heating more generally \citep[][and references therein]{Adhikari2019}.


For these reasons, crossing this boundary was a key design driver for the trajectory of Parker Solar Probe \citep[Parker;][]{Fox2016}. Prior to launch, and during the early phases of the mission, this had to be estimated without direct confirmation of the ground truth either from radial scaling \citep{Kasper2019,Liu2021, Verscharen2021,Cranmer2023} or global coronal modeling \citep[e.g. ][]{Chhiber2022}. These approaches produced broadly consistent estimates in the range of 10\,--\,20\,R$_\odot$ \citep{Cranmer2023} with some Solar\,Cycle dependence \citep{Katsikas2010, Goelzer2014,Kasper2019}, sufficient to provide mission design constraints, and this was vindicated by a first crossing in 2021 \citep{Kasper2021} at 19.8\,R$_\odot$. Since then, Parker has continued to dive routinely below the Alfv\'en surface and has now built sufficient statistics to provide this previously missing ground truth over half of the Solar\,Cycle. Moreover, the constraints not only determine the radial distance but also the spatial variation in the structure \citep{Liu2023,Badman2023,Finley2025} across significant portions of the corona sampled over just a few days.  

\citet{Finley2025}, in particular, has recently used this spatial sampling from Parker, and a similar but simpler scaling method to what is explored in this work, to derive geometrical constraints of the portion of the Alfv\'en surface sampled during Encounters 4-20. They observe a general increase in surface height and angular momentum loss with solar cycle consistent with prior work \citep[e.g. ][]{Katsikas2010,Goelzer2014} and relate features of the topology to coronal magnetic topology. The results of this work are highly consistent with the results presented here and will be useful to cross-reference further in this paper.

In this work, we leverage these new powerful constraints and coincident synoptic inner heliospheric measurements from Solar Orbiter \citep{Muller2020} and from multiple spacecraft at L1 (Wind, \citet{Wilson2021}; the Advance Composition Explorer ACE, \citet{Stone1998}; and the Deep Space Climate Observatory DSCOVR, \citet{Burt2012}), to systematically map the solar Alfv\'en surface's morphology over the ascending phase and maximum of Solar\,Cycle\,25. We present the scaling methodology (Section\,\ref{sec:methods}), ground truth comparison and validation (Section\,\ref{subsec:validation}) and then present the resulting determined structure as 2D near-equatorial plane cuts (Section\,\ref{subsec:2d}), distributions of heights (Section\,\ref{subsec:height_vs_solcyc}) and departures from spherical symmetry (Section\,\ref{subsec:non-uniformity}). We close with implications of our results for placing our Sun in stellar context and interpreting different sub-Alfv\'enic intervals seen by Parker (Sections\,\ref{sec:discussion} and \ref{sec:conclusions}).   

\section{Methodology}\label{sec:methods}

To estimate the location of the Alfv\'en surface, we develop a family of physics-based radial scalings for how the solar wind radial bulk speed ($V_{SW}(R)$) and Alfv\'en speed ($V_A(R)$) evolve with heliocentric distance ($R$) which are sized to span observed statistical trends. We then choose pairs of profiles based on in situ measurements of $V_{SW}$ and $V_A$ at heliocentric distance $R$ and compute their intersection point distance and critical speed ($R_A,V_A(R_A)$). Lastly, we use Parker spiral backmapping \citep[e.g.,][]{Nolte1973} to associate the 3D measurement location (in spherical Carrington-frame coordinates, $R,\theta,\phi$) to a given 3D position of the Alfv\'en critical point associated with that measurement ($R_A,\theta_A,\phi_A$). We note that the Parker spiral has zero meridional flow such that $\theta_A = \theta$ and that for most of this work, we assume latitude can be neglected and that we predominantly examine structure we approximate as being in the solar equatorial plane (an assumption that will soon be able to be relaxed with Solar Orbiter's exploration of high latitudes). The full procedure may therefore be described by the mapping: 
\begin{align}
    \big[V_{SW}(R,\theta,\phi), V_A(R,\theta, \phi)\big] \rightarrow \big[R_A, \theta, \phi_A, V_A(R_A)\big]
\end{align}

In this section, we first briefly describe the methodological steps we take to prepare and develop the in situ data set and the radial scaling profiles with further details provided in the appendices. We then illustrate the intersection method and provide validation of the method using Parker ground truth measurements of the actual locations of Alfv\'en surface crossings.

\subsection{Dataset}\label{subsec:data-prep}

For this work, we require in situ measurements of the solar wind radial proton velocity ($V_{SW}$) and the solar wind radial Alfv\'en speed $V_A = B_R/\sqrt{\mu_0 m_p N_p}$\footnote{$\mu_0$ is the magnetic permeability of free space and $m_p$ is the proton mass.}, the latter of which, requires measurements of the solar wind magnetic field ($B_R$) and proton density ($N_p$). Additionally, to size the thermal pressures used in developing the radial profiles, we use the measured scalar proton temperature $T_p$.  We develop a uniform data set of all these quantities averaged over 15--minute intervals over the course of the first 23\,orbits of the Parker mission from October\,2018 through to April\,2025. Within this time-span, we compute these quantities wherever available as measured by Parker, Solar Orbiter and by multiple spacecraft at L1.

A full accounting of the data sources, coverage, and steps to combine measurements taken by different instruments (where needed) is provided in Appendix\,\ref{sec:appendix:data}. The outcome is a well-vetted dataset of $N_p, V_{SW}$, $T_p$, $B_R$, $V_A$, and $M_A = V_{SW}/V_A$ covering the time range mentioned above (10/2018--04/2025), which comprises the ascending phase and the peak of the Solar\,Cycle\,25, and with measurements spanning from 9.86\,$R_\odot$ (as of December\,2024) out to 1\,au ($\sim$215\,$R_\odot$).

\subsection{Generating Solar Wind Speed and Density Profiles}\label{subsec:isopoly-profiles}

Having produced the in situ dataset, we compute statistics as a function of heliocentric distances to accurately size radial profiles of the solar wind speed and Alfv\'en speed. The procedure we follow, fully laid out in Appendix\,\ref{sec:appendix:solar-wind-modeling}, is to use ``isopoly'' two-fluid solar wind models \citep{Dakeyo2022} which combine an isothermal coronal portion with proton and electron temperatures, $T_p$ and $T_e$ respectively and a polytropically cooling solar wind portion with different polytropic indices for protons and electrons {($\gamma_p$ and $\gamma_e$, respectively), with an interface height between the two regimes of $R_{iso}$.

We additionally provide an external empirically-sized force, F(R), (i.e., additional to thermal pressure gradients) which is required to produce acceleration profiles which match faster asymptotic speed winds and is generally attributed to Alfv\'enic fluctuations close to the Sun \citep{Shi2022, Rivera2024a, Rivera2025}. These models allow realistic acceleration profiles to be produced and to provide mass-flux conserving density profiles to be derived in turn. Informed by the bounds initially fit by \citet{Dakeyo2022} and the forcing function in \citet{Rivera2024a}, we size the force function to be progressively stronger for higher wind speeds at 1\,au.

The final outcome is a family of 40\,profiles of proton densities, proton velocity, proton temperature and electron temperature ($N_{p,i}(r),V_{SW,i}(r),T_{p,i}(r),T_{e,i}(r)$ respectively), where $i\in \{1,40\}$ indexes the different profiles and is ordered according to increasing solar wind speed at 1\,au. We reserve the lower-case symbol $r$ to mean heliocentric distance as an independent variable, as opposed to the $R$ coordinate where a given measurement is taken. 

These profiles, except for the electron temperature for which a robust in situ dataset is not currently readily available, are sized to span the 1st\,--\,99th percentile ranges of the dataset as a function of distance self consistently for solar wind density, proton speed and proton temperature, as illustrated in Appendix Figure\,\ref{fig:model-profiles}. Specifically, the proton parameters $T_p$ and $\gamma_p$, along with the interface height, $R_{iso}$, are set to match statistics of the radial evolution of $T_p$ in the solar wind.  Electron parameters ($T_e$ and $\gamma_e$) are determined based on prior in situ and remote sensing work \citep{Dakeyo2022, Rivera2025}. Finally, The external force, F(R) is then proportionally sized to match statistical solar wind velocity profiles. A full accounting of the model parameter ranges is given in Table \ref{table:appendix:parameter-val}.

\subsection{Alfv\'en Speed Profile}\label{subsec:alfvenspeed-profiles}

The final ingredient needed is to use the derived mass-flux conserving density profiles and some choice of normalization to produce an Alfv\'en speed profile corresponding to a given acceleration profile. Once this choice is made, the intersection location of the two profiles (or equivalently, the location where the ratio, $M_A=1$) follows directly. 

For a given in situ measurement of $N_p$, $V_{SW}$ and $V_A=B_R/\sqrt{\mu_0 m_p N_p}$ at heliocentric distance $R$, the procedure is to choose the $k^{th}$ velocity profile $V_{SW,k}(r)$ which most closely passes through the coordinate ($R$, $V_{SW}$) via nearest neighbor search. We then select the corresponding mass--flux conserving proton density profile $N_{p,k}(r)$ and compute:
\begin{align}
  V_{A,k}(r) = V_{A}\frac{R^2}{r^2}\sqrt{\frac{N_{p,k}(R)}{N_{p,k}(r)}}  = V_{A}\frac{R}{r}\sqrt{\frac{V_{SW,k}(r)}{V_{SW,k}(R)}}
\end{align}

which describes a radial Alfv\'en speed profile which:
\begin{itemize}
    \item Assumes magnetic flux conservation ($B_{R}(r) \propto 1/r^2$). 
    \item Enforces mass flux conservation given the associated acceleration profile.
    \item Matches the measured $V_A$ at measurement distance $R$.
\end{itemize}

We see that in the limit $\frac{V_{SW,k}(r)}{V_{SW,k}(R)} \rightarrow const $ (i.e., as the wind speed profile asymptotes), we recover the simpler approximation that $V_A(r) \propto (1/r)$ \citep{Liu2023}. 

\subsection{Alfv\'en Surface Intersection and Mean Behavior}\label{subsec:intersection-method}

\begin{figure*}
    \centering
    \includegraphics[width=0.9\linewidth]{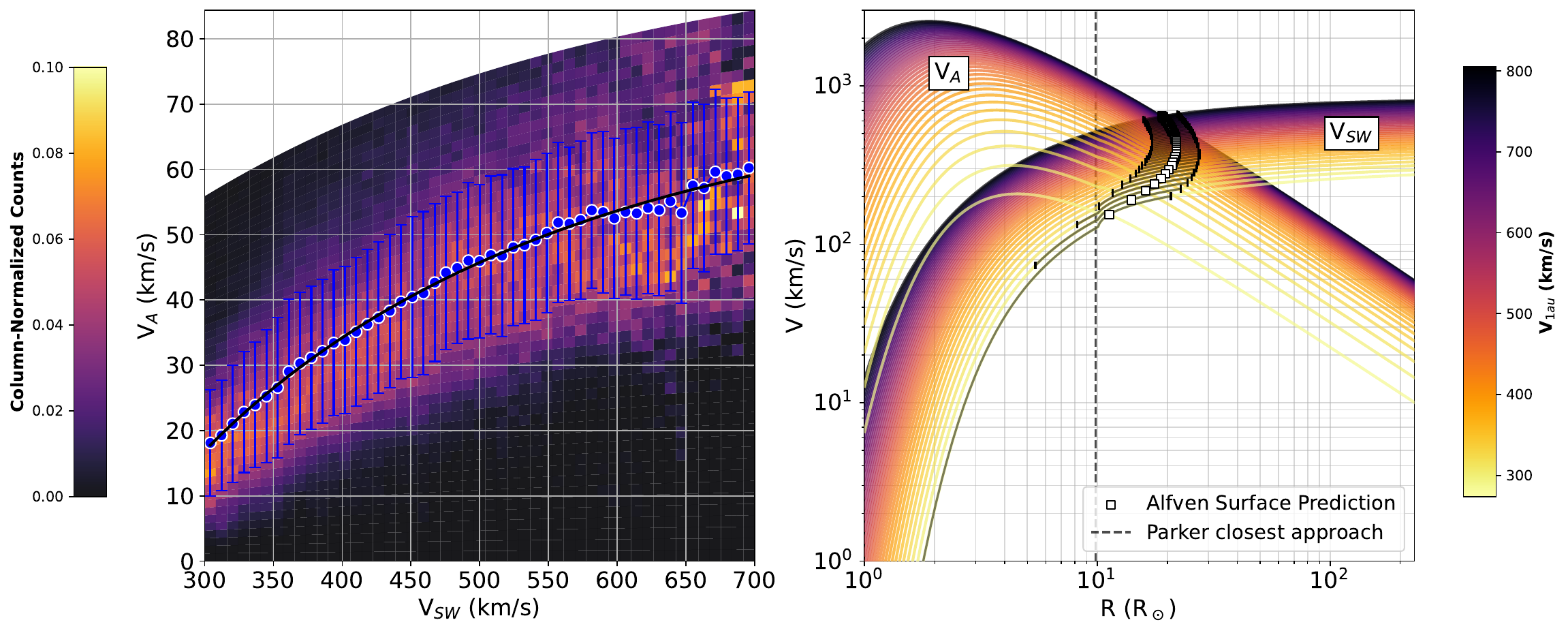}
    \caption{\textbf{Illustration of Scaling Intersection Method}. {\it Left} Panel: 2D Column-Normalized Histogram of V$_A$ and V$_{SW}$ observations at L1 with mean ({\it blue}), standard deviations ({\it blue} bars) and a fit to the mean ({\it black}) showing a clear monotonic relationship. {\it Right} Panel: ``Isopoly'' V$_{SW}$ and V$_A$ profiles (see Appendix\,\ref{sec:appendix:solar-wind-modeling}) colored by asymptotic wind speed reflecting the systematic relationship at 1\,au from {\it left} panel, and their intersections and ranges in V$_R$\,--\,R space. The propagated standard deviation of the intersection along each speed profile is indicated with {\it black} bars. In this work, we estimate the Alfv\'en surface height and critical speed with these types of intersections with the specific $V_{SW}$ and $V_A$ curve selected by the nearest--neighbors for a given measurement ($R,V_{SW},V_A$).
}
    \label{fig:intersection}
\end{figure*}

We now have all the ingredients required to compute an implied Alfv\'en surface location for an arbitrary in situ measurement of $V_{SW}$ and $V_A$ at distance $R$ and heliographic angular coordinates $\theta,\phi$.

To estimate the altitude of the Alfv\'en surface, we compute the intersection point, $R_A$, of $V_{SW,k}(r)$ and $V_{A,k}(r)$. This is illustrated in Figure\,\ref{fig:intersection} where we use the observed statistical correlation at L1 between solar wind speed and Alfv\'en speed ({\it left} panel) to choose a well organized set of Alfv\'en speed profiles. The resulting intersection and systematic interaction shape are shown in the {\it right} panel where the families of solar wind speed and Alfv\'en speed profiles are both shown colorized by their asymptotic solar wind speed at 1\,au \citep[similar to the method used to compute intersections heights for different solar wind speeds in ][]{Dakeyo2024a}.  Additionally, to communicate that the specific mass--flux normalization will vary based on the specific in situ measurements, we also connect error bars in the the L1 statistical correlation to resulting error bars in the intersection points along the $V_{SW}$ profile. This illustrates a general statistical expectation that faster speed wind will have a higher average Alfv\'en surface than slower speed winds, but also that there is significant spread. Note that this spread is larger than the sensitivity due to quantization of the wind profiles (the difference between neightboring white intersection points) which ranges from negligible for fast asymptotic speeds up to $\pm0.7 R_\odot$ for the slowest speeds.

Finally, to assign this Alfv\'en surface location to a 3D position, we connect the measurement longitude ($\phi$)  and latitude ($\theta$) to a corresponding longitude ($\phi_A$) and latitude ($\theta_A$) of the Alfv\'en surface location using a ballistic Parker spiral according to the measured wind speed $V_{SW}$ using equation (1) quoted in \citet{Badman2020}. In this step, we note that the latitude is limited primarily by the sampling latitude of the measurements and zero meridional flow or expansion is assumed explicitly using a Parker spiral such that $\theta_A=\theta$. The accuracy of the longitude, meanwhile is limited by the ballistic assumption with errors dependent on the starting point of the mapping and its velocity \citep{Dakeyo2024a}.  

After this final step, we have obtained a mapping for any given measurement ($R,\phi,\theta, V_{SW}, V_A$) to an Alfv\'en surface location ($R_A,\phi_A,\theta_A$).

In the remainder of the paper, we will discuss and present the outcomes, accuracy and physical interpretation of the result of applying this mapping to the full dataset described above. 

\section{Validation}\label{subsec:validation}


\begin{figure*}
    \centering
    \includegraphics[width=0.9\linewidth]{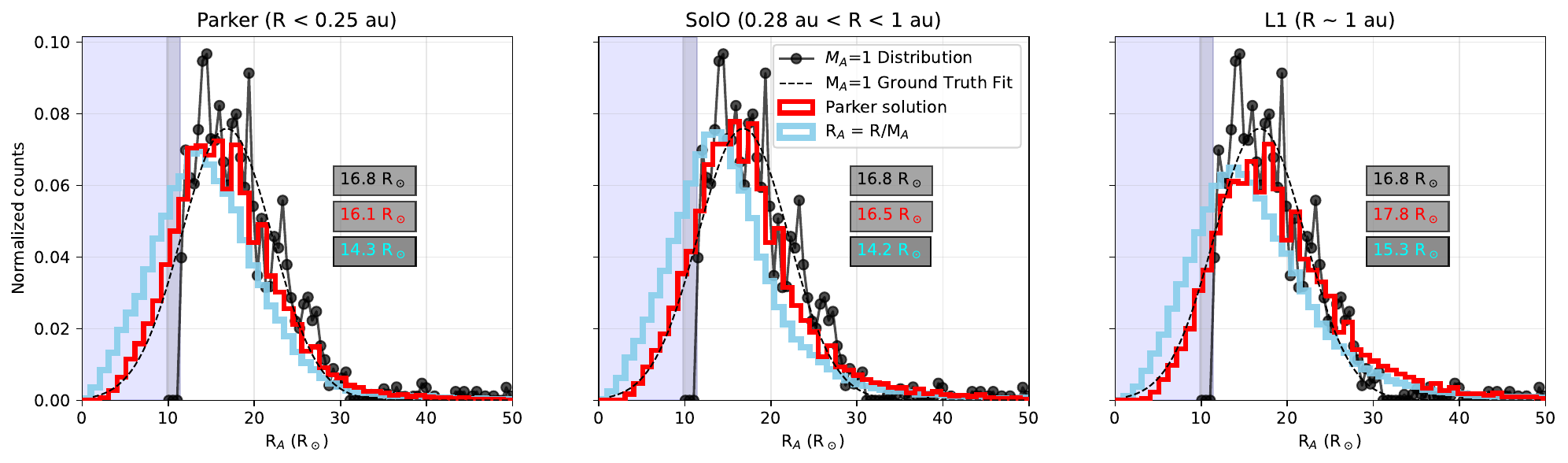}
    \caption{\textbf{Validation of the scaling method}. Each panel shows the normalized ground truth distribution of the heights of measured Alfv\'en surface crossings ({\it black} solid line and dots) and a gaussian fit to this distribution ({\it black} dashed line). Each panel also includes {\it red} and {\it cyan} distributions which show the results respectively of the isopoly intersection method and the simpler construction R$_A$=R/M$_A$  as computed (from left to right) from Parker data within its encounters ($R < 0.25$\,au), Solar Orbiter from 0.28\,--\,1\,au and from all observations at L1 over the total time period examined. A shaded {\it blue} region indicates the ranges of distances in the corona not probed by Parker, and the {\it light gray} shading indicates the narrow range of distances probed in Parker’s most recent encounters. Inset numbers show the distribution median with colors corresponding to respective distributions ({\it black} - ground truth fit, {\it red} - Parker Wind scaling, {\it cyan} - R/$M_A$ scaling}
    \label{fig:methods:validation}
\end{figure*}

To validate the estimated height of the Alfv\'en surface, we use the unique capability of Parker to directly encounter the Alfv\'en surface and therefore provide a ground truth distribution of its location. Here, we define any measurement in our 15--minute binned data set in which $0.95 < M_A < 1.05$ as such an encounter. Recording all such instances, we build up a distribution of the altitude of Parker for each measurement within equally spaced 1\,$R_\odot$ bins. Next, to account for the varying dwell time spent by the spacecraft at different heliocentric distances, we normalize this distribution by the number of 15--minute intervals spent in each of these heliocentric distance bins which ranges from a minimum of 70 intervals for innermost bin up to several hundred for further heliocentric distances. 

In Figure\,\ref{fig:methods:validation}, we present this distribution in {\it  black} solid line and dots, and further fit a gaussian to it in {\it black}  dashed lines (which also provides a normalization for the raw distribution, accounting for the altitudes below 9.86\,$R_\odot$ which Parker does not sample). In the three columns of the figure, in {\it red}, we overlay distributions from scaling the full dataset for Parker encounters (that is, measurements for which $R < 0.25$\,au), Solar Orbiter (ranging from 0.3\,--\,1\,au) and from L1 spacecraft (at $\sim1$\,au) respectively. Additionally, in {\it cyan}, we overlay the distribution from simple scaling without accounting for acceleration and the knock--on effect on mass--flux. In both cases, since our hypothesis is that the resulting distribution should not depend on what distance the scaling is started from, we do not need to perform further normalization other than dividing by the total number of counts. 

Immediately, we see that the scaled distributions ({\it red} curves) match the ground truth very closely and the distribution from each of Parker, Solar Orbiter and L1 all give very similar results, with the exception that a high--$R_A$ tail is apparent in the Solar Orbiter and L1 results, and appears to get larger with distance. This means the scaling we are using is approximately independent of radius which validates that our profiles represent the observed statistical trends of radial evolution well. In contrast, the {\it cyan} curves whose profiles neglect the solar wind acceleration and mass--flux conservation show worse agreement with the ground truth distribution is in all cases. To quantify this agreement, we compute the median statistic of each distribution and display them in Figure\,\ref{fig:methods:validation}. This shows that in all cases, neglecting acceleration results in a distribution of Alfv\'en surface heights which is an underestimate (by 1.5\,--\,2.5\,$R_\odot$). Accounting for solar wind evolution, the distribution medians are much closer together with differences ranging from 0.3\,--\,1.0\,$R_\odot$. 

We conclude that our scaling method accurately reproduces the observed distribution of Alfv\'en surface crossings from Parker, and is an improvement relative to scaling approaches which neglect solar wind acceleration \citep[e.g.,][]{LiuY2021,Liu2023,Badman2023} which produce systematic underestimates.

\section{Results}\label{sec:results}

Next, we present the inferred geometry of the Alfv\'en surface over the time interval spanning the rising phase and just past the peak of Solar\,Cycle\,25. In these following sections, we illustrate its overall 2D shape (Section\,\ref{subsec:2d}, its average height (Section\,\ref{subsec:height_vs_solcyc}) and lastly, its deviation from spherical symmetry and its roughness (Section\,\ref{subsec:non-uniformity}). In these sections, we separate the data by the 23\,Parker orbits that occurred over this time interval.

\subsection{2D Geometry}\label{subsec:2d}

\begin{figure*}
    \centering
    \includegraphics[width=0.7\linewidth]{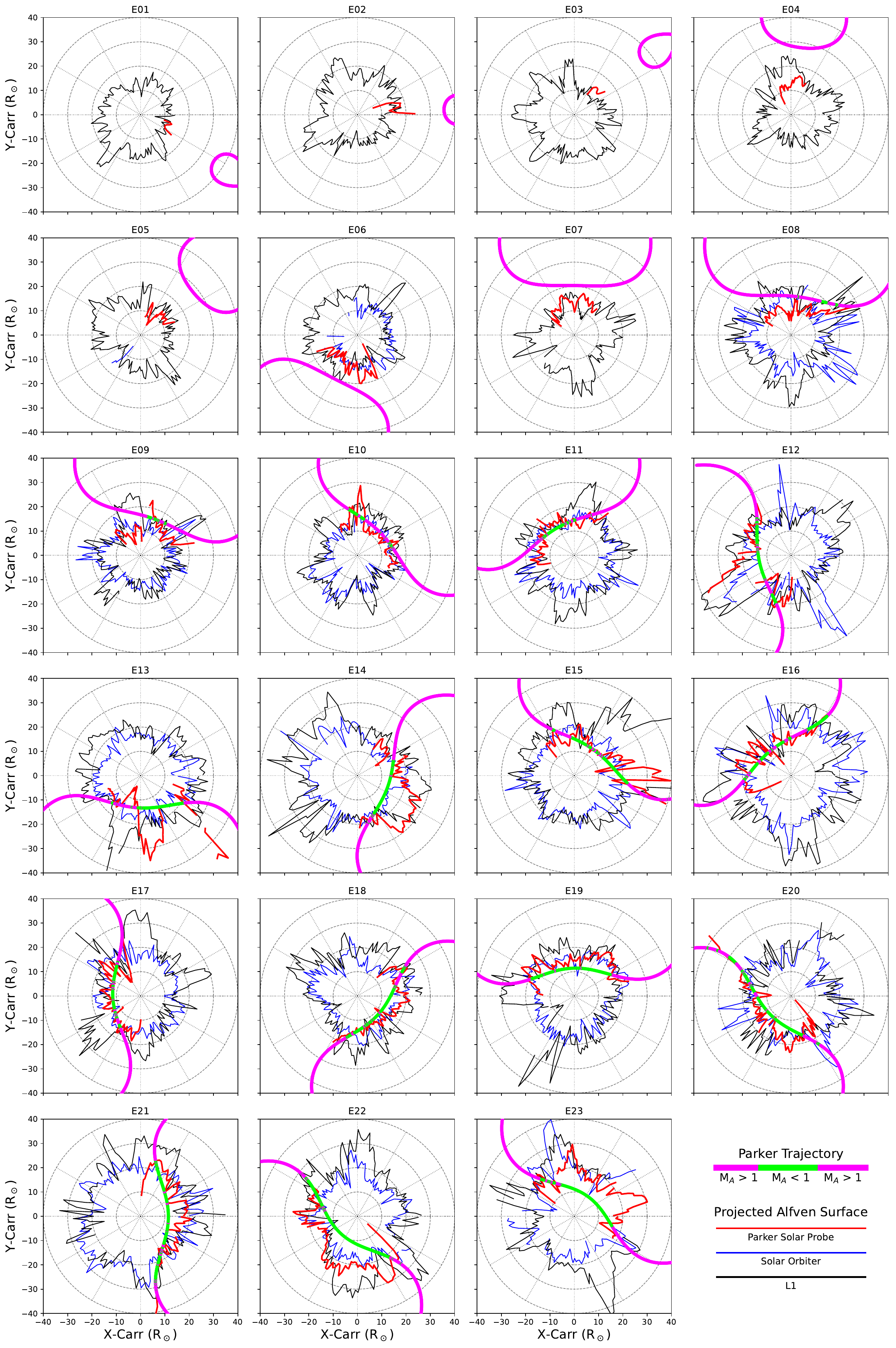}
    \caption{\textbf{The Shape of the Solar Alfv\'en Surface 2018--2025}. Each panel shows the inferred 2D shape of the Sun’s Alfv\'en surface near the solar equatorial plane as scaled from L1 data ({\it black}), Solar Orbiter data ({\it blue}) and Parker data ({\it red}) plotted in the Carrington frame. From {\it top left} to {\it bottom right}, the plots advance chronologically by Parker encounter. The orbit of Parker in the Carrington frame is shown in each case and is colored according to whether it measured super- ({\it magenta}) or sub- ({\it lime}) Alfv\'enic wind.}
    \label{fig:alfven-surface-topdown}
\end{figure*}


In Figure\,\ref{fig:alfven-surface-topdown}, we plot multi-spacecraft measurements of the 2D geometry (in the Carrington frame, i.e. in coordinates which co-rotate with the Sun) of the Alfv\'en surface for ascending and peak phases of Solar\,Cycle\,25, projected onto the solar equatorial plane (i.e. neglecting variation in latitude). We use available measurements from Parker, Solar Orbiter and L1, divided up in time according to each Parker orbit. For Parker, we use encounter data only ($R < 0.25$\,au) and get an instantaneous cut over a portion of a Carrington rotation which is very small at the start of the mission but approaches ~170$^\circ$ by the most recent orbits. Meanwhile the further out data from Solar Orbiter and L1 can provide full 360$^\circ$ longitudinal coverage. For each of these measurement points, we take measurements over one full rotation around the Sun, centered on the date of Parker perihelion. For the L1 measurements, this means taking a period of time equal to a Carrington rotation ($\sim$27\,days), while for Solar Orbiter the appropriate length of time varies based on the spacecraft's distance from the Sun with a full rotation taking longer at its $\sim$0.3\,au perihelia (up to $\sim$50\,days). The results are shown in Figure\,\ref{fig:alfven-surface-topdown}.

For each spacecraft, we take the Carrington longitude and inferred altitude of the Alfv\'en surface location from the full 15--minute dataset and bin the results into 1--degree segments of longitude. For each bin, we compute the median altitude. The resulting surfaces computed from Parker are plotted in {\it red} and are directly comparable to the partial cuts shown by \citet{Finley2025} in their Figures 2 and C.1. Data from Solar Orbiter is plotted in {\it blue} and from L1 in {\it black}, and this color-scheme is followed in Figures\,\ref{fig:height-solar-cycle} and \ref{fig:roughness}. Solar Orbiter data is only available from Parker Orbit\,5 and higher due to the later launch of the mission (January\,2020). 

The trajectory of Parker in each time interval is also plotted in each panel of Figure\,\ref{fig:alfven-surface-topdown} and colored {\it magenta} for super-Alfv\'enic intervals and {\it lime} for sub-Alfv\'enic intervals. This clearly demonstrates how the direct Parker observations of crossing the Alfv\'en surface are highly consistent with not only the inferred Alfv\'en surface structure inferred by Parker (by construction) but also by the independent estimates of the surface from much further away with Solar Orbiter and L1. 

Comparing the panels from start to finish, we see the onset of Parker starting to sample sub-Alfv\'enic wind from Encounter\,8 onwards \citep{Kasper2021} driven not just by the dropping perihelion distance but also by an increase in the average height of the Alfv\'en surface. This increasing altitude trend continues as the Solar\,Cycle has progressed to date, with the last panel (Encounter\,23) showing the most recent Parker encounter which occurred in full solar maximum. Simultaneous to the perihelion distance dropping, the Parker orbits increasingly complete longitudinal coverage is clearly shown, with the most recent orbits spanning almost halfway around the Sun in Carrington longitude. 

We see compelling evidence that the large scale evolution of this 2D structure is real given its similar structure at all stages of evolution as inferred from scaling far from the and as measured directly by Parker, i.e., confirmed by independent asynchronous data. We also see instances of an anomalously high altitude which lies beyond the average height but is very localized in longitude. Especially when these extrusions occur in one spacecraft's inferred surface but not others, these occurrences appear to be related to large CMEs (with direct evidence at least for Encounters\,10 \citet{McComas2023,Jagarlamudi2025} and 13 \citet{Romeo2023}) that Parker crosses behind. More broadly, these instances may represent a more general class of transient wake solar wind, the exploration of which will be a follow up paper. 

Lastly, in the last two panels in Encounters\,22 and 23, we observe the deepest crossing into sub-Alfv\'enic wind by Parker to date in its first record-breaking orbit at 9.86\,$R_\odot$. While in prior close approaches, the trajectory has generally been observed to skim the inner boundary of the Alfv\'en surface, the trajectories from Encounters\,22 and 23 show, for the first time, a clear set of measurements far from any rough boundary effects and provides an important test dataset for differing energization processes in the corona vs. the super-Alfv\'enic solar wind \citep{Ruffolo2024}. This situation is particularly well supported in that the average height of the Alfv\'en surface for this interval (around 20\,$R_\odot$) is estimated to be in the same place through our scaling both from out at 1\,au ({\it black}) and from deep inside it providing a good further cross validation of the applicability of the radial profiles used in this work. Further, the transition across the Alfv\'en surface for these orbits appears to be driven by radial motion of the spacecraft, rather than skirting along a rough boundary.

\subsection{Average Height}\label{subsec:height_vs_solcyc}

\begin{figure*}
    \centering
    \includegraphics[width=0.8\linewidth]{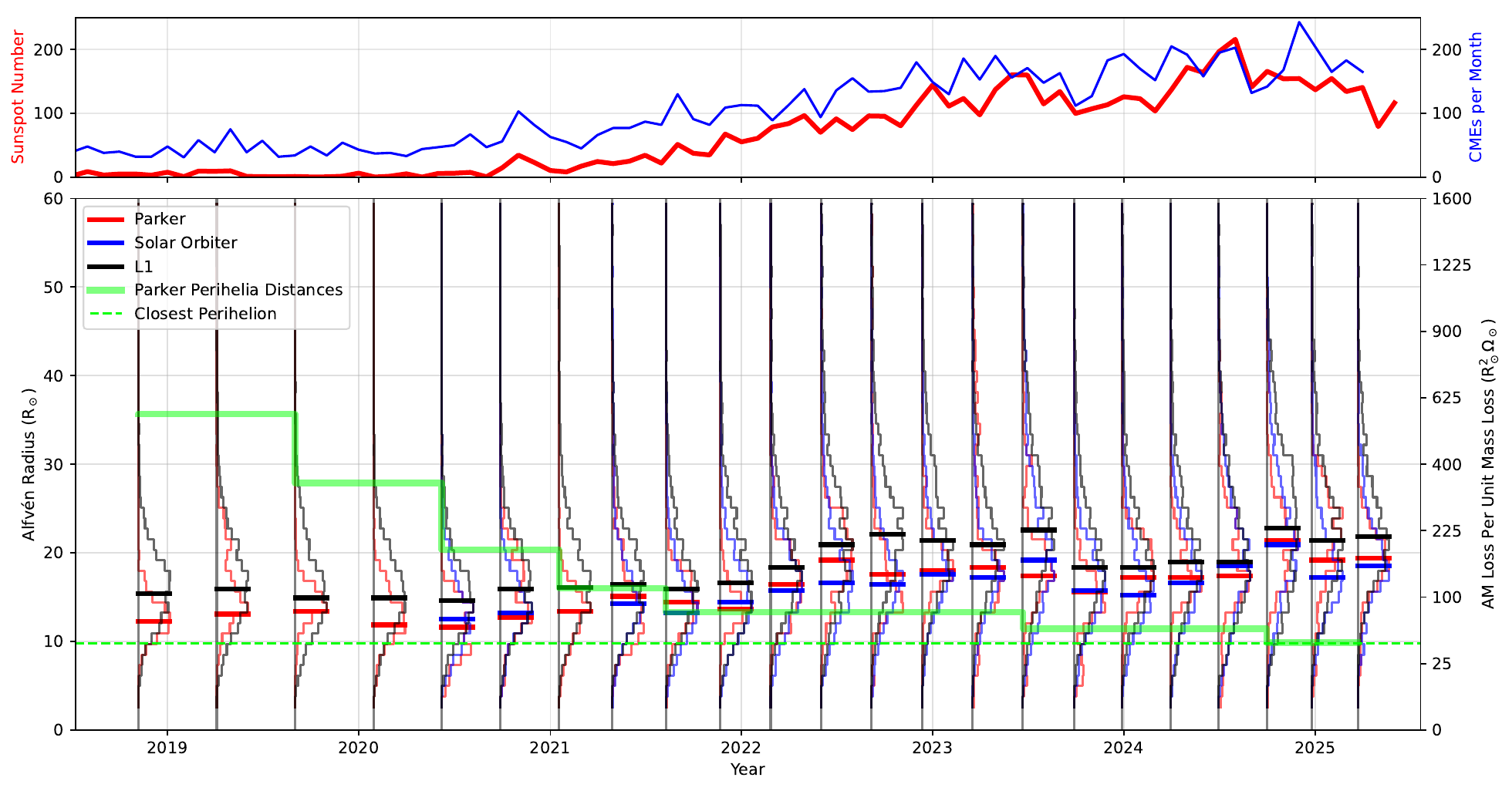}
    \caption{{\bf Distribution of Heights vs the Solar\,Cycle}. The main ({\it bottom}) panel shows the distribution of Alfv\'en surface heights oriented vertically with means indicated by short horizontal bars and colorized as in Figure\,\ref{fig:alfven-surface-topdown} ({\it black}--L1, {\it blue}--Solar Orbiter and {\it red}--Parker missions) and plotted a function of time with one distribution per Parker encounter. The {\it top} panel shows the monthly smoothed numbers of sunspots (in {\it red}, from \url{https://www.sidc.be/SILSO/datafiles}) and coronal mass ejections (in {\it blue}, from the SOHO/LASCO CME catalog) plotted on the same x-axis. A second y-axis on the {\it right} of the main panel computes R$_A^2 \Omega_\odot$, a figure of merit for angular momentum loss \citep[per unit mass loss, see][]{Weber1967}.
}
    \label{fig:height-solar-cycle}
\end{figure*}

Next, we collapse these 2D structures into 1D to quantify the Solar\,Cycle dependence of various properties, starting with the average altitude. 

In Figure\,\ref{fig:height-solar-cycle} we plot for each Parker orbit, a histogram of inferred heights independently produced from all three spacecraft (where available), following the same color scheme as used in the previous section. The median of each distribution is also shown as a horizontal bar in each case. The red bars and distributions corresponding to Parker medians are again directly comparable to \citet{Finley2025}'s Figure 3, and appear highly consistent. Each distribution is plotted along an x-axis conveying time. A thick, transparent {\it green} line shows the progression of Parker's perihelion distance with each orbit and a {\it green} dashed horizontal line highlights the closest perihelion distance of 9.86\,$R_\odot$ for comparison to the Alfv\'en surface height in earlier orbits. In the {\it top} panel, we plot the advancement of the Solar\,Cycle as communicated by the monthly smoothed sunspot number in {\it red} \footnote{Obtained from \url{https://www.sidc.be/SILSO/datafiles}}, and from the number of CMEs per month reported in the SOHO/LASCO CME catalog \citep{2008AnGeo..26.3103Y, 2009EM&P..104..295G, Gopalswamy2024arxiv} in {\it blue}. These profiles both illustrate that the range of time explored to date in the era of the Parker mission spans the first half of the Solar\,Cycle\,25. 

The average height and overall distribution of height of the surface is shown to increase in lockstep with the Solar\,Cycle indexed by both sunspots and CME counts, as expected \citep{Katsikas2010,Goelzer2014, Kasper2019, Finley2025}. This includes not just the long term monotonic change in solar cycle phase, but also appears borne out by a decrease in both sunspot number and average height from mid 2023 to early 2024.

The median is in the range 12\,--\,17\,$R_\odot$ at solar minimum and is more in the range of 15\,--\,23\,$R_\odot$ now at solar maximum. The lower limit of these ranges is very close to the 11-16 $R_\odot$ range determined by \citet{Finley2025}. The average inferred by the different spacecraft shows a dispersion of around 3\,--\,5\,$R_\odot$, around the order of the distribution standard deviations. The median height inferred from L1 measurements is always higher than the Parker and Solar Orbiter measurements, which are both typically separated by less than 1\,$R_\odot$. This is an interesting result as Solar Orbiter's heliocentric distance varies between 0.3 and 1\,au. If this effect was simply an issue with radial scaling, one would expect the Solar Orbiter median height to match the L1 height sometimes and Parker other times, but it always lies closer to Parker. In any case, this observed dispersion between different spacecraft provides a conservative estimate of errors associated with our method, and is potentially related to recent work advocating for an Alfv\'en ``region'' rather than a strictly 2D ``surface'' \citep{Chhiber2022,Chhiber2024}. Further, the distributions in Figure\,\ref{fig:height-solar-cycle} also clearly exhibit an increasing standard deviation with solar activity across all spacecraft. More details on this finite thickness and asphericity are presented in the next section.



\subsection{Asphericity and Roughness}\label{subsec:non-uniformity}

\begin{figure*}
    \centering
    \includegraphics[width=0.9\linewidth]{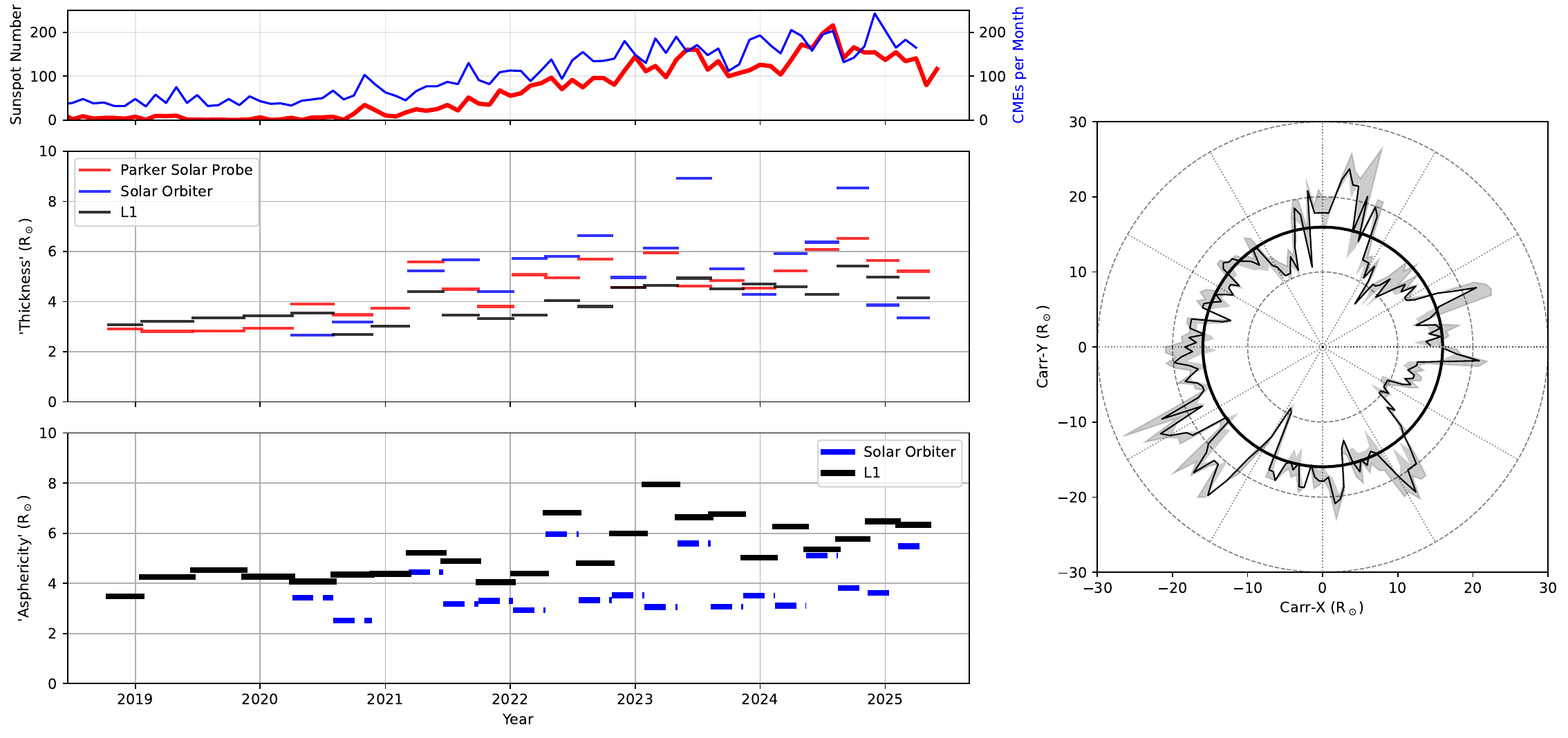}
    \caption{\textbf{Thickness and Asphericity of the Alfv\'en Surface}. As in Figure\,\ref{fig:height-solar-cycle}, geometric properties of the Alfv\'en surface as a function of time are compared to the monthly smoothed number of sunspots (shown in {\it red}) and coronal mass ejections (in {\it blue}). In the {\it middle left} panel, the average ``thickness'' of the surface is plotted vs. time while the {\it bottom left} panel shows the ``asphericity'', both in units of solar radii and computed from different spacecraft with the same color scheme as in Figures\,4 and 5. The inset on the right illustrates the meaning of these two quantities showing the Encounter\,1 Alfv\'en surface computed from L1 data. In this plot, the {\it black} noisy line is the average shape as shown in Figure\,\ref{fig:height-solar-cycle}, while the {\it gray} region is the standard deviation of computed altitudes for each longitude bin. The ``thickness'' is then the average of this standard deviation over all longitudes. The asphericity is defined via the mean difference between the mean shape of the surface ({\it black} noisy line) and longitude-averaged height ({\it black} circle).}
    \label{fig:roughness}
\end{figure*}

In Figure\,\ref{fig:roughness}, we present two quantitative measures of irregularity of the Alfv\'en surface structure as a function of time: 

\begin{itemize}
    \item A ``roughness'' parameter which is the standard deviation of the height in each longitude bin. 
    \item An ``asphericity parameter'' which is the standard deviation of the 2D surface shown in Figure\,\ref{fig:alfven-surface-topdown}, i.e. its deviation from spherical symmetry in one degree bins.
\end{itemize}

These two quantities are illustrated further in the {\it right} panel of Figure\,\ref{fig:roughness} which shows the Alfv\'en surface for Parker Encounter\,1 including its 2D median contour, its median height as a {\it black} circle, and the `thickness' as a {\it gray} region in 2D.  The thickness then is the average width of the {\it gray} region, while the asphericity is the extent to which the {\it black} solid curve deviates from the median circle. These may be regarded as the large and small scale limits to the power spectra presented by \citet{Finley2025} in their Figure 5. 

Again, the same quantities are computed independently by the three points of measurement and follow the same color scheme as previously. In the {\it top} panel, the monthly smoothed sunspot number and CME rate is repeated for reference. The two lower panels then show respectively the thickness, and the asphericity. The asphericity as computed by Parker is omitted due to it being strongly systematically distorted by the range of longitude it probes at perihelion which may explain the inconsistency with \citet{Finley2025} Figure 5 in this case which only explores the partial Alfv\'en surface cuts of Parker.

As with the average height, the thickness is clearly correlated with solar activity and shows similar values across all measurements. For asphericity, the L1 data shows a convincing trend of being closer to spherical at solar minimum and then a transition to a less spherical and and more rapidly changing state at solar maximum. Solar Orbiter data supports this inference to a limited extent due to the lack of measurements during true solar minimum conditions, but does suggest a large variability in asphericity at solar maximum.


\section{Discussion}\label{sec:discussion}

In this work, we present new measurements of the evolving time-dependent structure of the Sun's Alfv\'en surface using novel new near-Sun measurements from Parker Solar Probe and Solar Orbiter, as well as synoptic observations from L1. We leverage these datasets' unprecedented combined coverage of basic solar wind parameters as a function of radial distance, including ground truth measurements of the Alfv\'en surface's true location provided by direct crossings from Parker, to produce a set of solar wind radial profiles which span the observed statistical data set and describe how the measurements taken at different radial distances transform into each other (Appendix Figure\,\ref{fig:model-profiles}). These profiles allowed us to obtain a mapping of a single point measurement of solar wind speed and Alfv\'en speed to an estimate of the 3D location of the Alfv\'en surface related to the plasma parcel of that measurement (Section\,\ref{subsec:intersection-method}).  

We validated this estimation scheme via comparing the inferred distribution of surface locations to the ground truth uncovered by Parker over the time it has been measuring sub-Alfv\'enic wind from 2021 to present (Figure\,\ref{fig:methods:validation}). This showed the simple radial profiles used here were a substantial improvement over approaches which neglect solar wind acceleration. Specifically, accounting for this increases the average altitude of the Alfv\'en surface height by approximately 3\,--\,4\,$R_\odot$.

This self-consistency additionally provides strong evidence that these profiles accurately describe the solar wind acceleration and mass--flux profile given the distributions were near identical when produced close to the Alfv\'en surface, all the way to 1\,au. Notably, these profiles not only provide physically justified temperature evolution in the outer corona and solar wind, but also incorporate recent findings of the increasing importance of Alfv\'enic fluctuation energy in achieving the upper end of the distribution of wind speeds observed at 1\,au \citep{Halekas2023, Rivera2024a}.

\subsection{Implications of 2D Geometry}
Having validated the scaling method, we proceeded to probe the implied Alfv\'en surface geometry from 2018 to 2025 (the first half of the Solar\,Cycle\,25 spanning from near solar minimum to past solar maximum). In Figure\,\ref{fig:alfven-surface-topdown}, we presented the 2D shape of the Alfv\'en surface in the solar equatorial plane at the time of each of Parker's first 23 encounters.  Independent estimates from Parker, Solar Orbiter and L1 provided cross--validation with close agreement identifying more robust, long--lived structure. Conversely, strong disagreement most often arises when one or more spacecraft exhibit a sudden outward extrusion of the Alfv\'en surface not observed in others. This likely points to transient disruptions to the steady solar wind picture assumed for scaling. Substantial inward extrusions in one spacecraft but not others is rare, with all spacecraft generally suggesting a consistent inner boundary for the Alfv\'en surface; Parker sub-Alfv\'enic intervals ({\it lime} intervals of the plotted trajectories) strongly support this and suggest the most recent encounters constitute true sub-Alfv\'enic flows.

Surveying from one encounter to another, the 2D extrapolations show a spiky, often aspherical surface which steadily inflates over time. Sudden, large extrusions in the surface also become more frequent in later encounters, also suggesting a connection to transients which occur more frequently with the Solar\,Cycle. 

\subsection{Average Height and Physical Implications}

Next, in Figure\,\ref{fig:height-solar-cycle}, we condensed this 2D structure into 1D distributions of the height and plotted the evolution vs. time and demonstrated the Solar\,Cycle evolution in this same period. This clearly demonstrated a monotonic relationship between the Solar\,Cycle and Alfv\'en surface height, not only of the median height but also for the overall distribution. This qualitative behavior was true for all spacecraft extrapolations, albeit with some dispersion (around 5\,$R_\odot$), and is consistent with prior work from a variety of methods \citep{Katsikas2010,Goelzer2014,Kasper2019, Finley2025}. A general systematically higher mean is observed between for L1 extrapolation and the other measurements at all times. This reason for this is somewhat unclear since Solar Orbiter is also sometimes at L1 and would be expected to show the same systematic behavior if it is an issue with radial scaling. The most likely difference is the use of multiple different instruments at 1~au, but needs further investigation.

Beyond just a rise from solar minimum to a peak at solar maximum, the median height shows evidence of a dip in height from mid 2023 to early 2024, coincident with a small dip in sunspot number, suggesting a connection between the Alfv\'en surface height and the magnetic structure of the corona (see \citet{Finley2025} for further discussion of this connection). 

On the same figure, we plotted a second y-axis converting the heights to R$_A^2 \Omega_\odot$ which is a figure of merit for angular momentum loss \citep[per unit mass flux, ][]{Weber1967}. \citep{Finley2025} shows that the longitudinally averaged mass flux is flat or even weakly decreasing with solar cycle, so this expression is strongly related to the true angular momentum loss rate. Owing to its quadratic dependence on $R_A$, this indicates that while the median height of the Alfv\'en surface increases only by $\sim$30\%, the rate of angular momentum loss approximately doubles. Additionally, since Solar\,Cycle\,25 is relatively modest in terms of peak sunspot number, this secular variation in angular momentum loss can be even more significant in strong Solar\,Cycles \citep[as has been noted through direct in situ measurements of torques in the solar wind][]{Finley2019},  and plausibly also on stars with more activity than the Sun. In any case, this strong variation stresses the need to account for secular solar--cycle behavior when placing the Sun in the context of stellar spin down rates and computing its overall spin--down lifetime \citep[e.g.][]{Chhiber2025}. 

A further physical implication for a varying median Alfv\'en surface height relates to the physics of coronal heating: It has been suggested that the location of the Alfv\'en surface is connected to regions of preferential minor ion heating \citep{Kasper2019} and potentially also more broadly to turbulent heating \citep{Chen2020,Ruffolo2024}. Even in the case where the physics on either side of the critical point does not change stepwise, it remains a critical point describing the scale height of solar wind radial variation. Thus, it is also a figure of merit for the volume in which coronal heating occurs.  

The asphericity and average altitude of the surface itself can also be used as a constrain in coronal modeling more broadly to test different coronal heating mechanisms. Specifically, the physics required to produce realistic Alfv\'en surface structures from a given magnetogram can be tested. For example, this may be able to distinguish between models in which Alfv\'en wave dissipation in the chromosphere is the primary heating mechanism \citep{vanderHolst.etal:10} as opposed to, for example, more impulsive mechanisms \citep{Chitta2023,Rawafi2023}.

\subsection{Implications of Departures from Spherical Symmetry}

In Figure\,\ref{fig:roughness}, we quantified the shape of the Alfv\'en surface beyond its median height via two quantities: the standard deviation in the computed height over all longitudes (which we termed the ``thickness''), and the overall deviation of the 2D median surface from its overall median height (which we termed the ``asphericity''). 

We computed these quantities and again cross--compared to the Solar\,Cycle which revealed the Sun's Alfv\'en surface is in general spikier or more variable at a small scale and also more deformed from spherical symmetry with increasing the Solar\,Cycle. These phenomena are likely related to both the more complex magnetic structure near the Sun's equator at solar maximum as well as the increased prevalence of eruptive and transient structures (as shown by the increase in monthly CME count plotted in Figures\,\ref{fig:height-solar-cycle} and \ref{fig:roughness}). Such structures are frequently identifiable in e.g. Figure\,\ref{fig:alfven-surface-topdown} as they appear in one spacecraft and not the others, and appear as sudden outward extrusions in the apparent height owing to the relative decrease in Alfv\'en speed compared to neighboring solar wind at the same heliocentric distance. This has the further implication that the early sub-Alfv\'enic crossings by Parker Solar Probe in which such outwards extrusions were observed \citep[][]{Kasper2021,Badman2023} could well be related to localized transients. There is some direct evidence of this for the outwards extrusions seen specifically in Encounter\,10 in \citet{Badman2023} as reported by \citet{McComas2023} and \citet{Jagarlamudi2025}. The fact these outwards extrusions can be inferred both by measurements far from the Sun and from crossing behind CMEs close to the Sun even well after the CME has passed is notable: It suggests that, although CMEs are strictly a transient phenomenon, they may in fact leave behind a long lived, steady low-Alfven mach number wake which can still be understood as a steadily evolving stream. Examining this more concretely will be the subject of follow up work.

Beyond CMEs, it is also possible some could be related to the compression regions of stream interaction regions which also correspond to relative increases in Alfv\'en speed and mach number \citep[e.g.][]{Dumbovic2022}, resulting in spikes in Alfv\'en surface height.

We note the general qualitative trend of a more variable surface with increasing solar cycle is different to what is reported by \citet{Finley2025} who observes a weak or even slightly anti-correlated relationship with solar activity levels over a range of spatial scales. This discrepancy likely relates the changing longitudinal sampling of Parker over the mission, which was the reason we didn't compute asphericity for Parker in Figure \ref{fig:roughness}, but may also be related to increased likelihood of CME detection for measurements further from the Sun which dwell at a given longitudes for longer times.



\subsection{Parker's Journey into Steady Sub-Alfv\'enic Wind}

Combining the 2D surfaces (Figure\,\ref{fig:alfven-surface-topdown}) and height distributions (Figure\,\ref{fig:height-solar-cycle}) with the Parker trajectories, the history of Parker's sub-Alfv\'enic measurements is clearly explained: Over the course of the prime mission, Parker's decreasing perihelion distance \textit{and} the increasing average altitude of the Alfv\'en surface have conspired to make sub-Alfv\'enic crossings increasingly likely. 

Early in the mission, the average height was relatively low, and Parker's perihelia remained well above it. Starting in Encounter\,8, Parker began to clip the top of the Alfv\'en surface, typically crossing outward extrusions in longitude as opposed to diving below it through radial motion. As the Alfv\'en surface continued to balloon, and Parker's perihelion approached its closest approaches, these crossings gradually changed to skimming the inner boundary (Encounters\,17\,--\,21). 

Finally in the most recent orbits (Encounters\,22 and 23), these crossings became unambiguous radial scans entering the region deep below the Alfv\'en surface. This suggests these and subsequent orbits are key for probing outstanding questions about whether heating or turbulence physics differs above and below the critical surface. 

The {\it red}  dashed curve in Figure\,\ref{fig:height-solar-cycle} also shows that at Parker's current perihelion distance of 9.86\,$R_\odot$, it is overwhelmingly likely to continue to probe well below the Alfv\'en surface even as solar activity declines into the next solar minimum and the average height correspondingly shrinks. Quantitatively, based on the solar minimum histograms plotted in Figure\,\ref{fig:height-solar-cycle}, at 9.9$R_\odot$, 98\% of predicted Alfv\'en surface locations are expected to be further from the Sun than Parker.

\subsection{Implications for stellar wind modeling and star-planet interaction  of other stars}

Constraints on departures from spherical symmetry and solar cycle dependence of the Sun's Alfv\'en surface may also provide useful constraints for modeling of stellar winds in general as well as for exoplanetary interactions.

Stellar wind modeling frameworks are developed primarily in the solar context with abundantly well observed boundary conditions\citep[e.g., the Space Weather Modeling Framework, SWMF][]{vanderHolst.etal:14,Gombosi.etal:18} and ways to validate directly with remote and in-situ data \citep{Sachdeva.etal:19, vanderHolst.etal:19}. However, when applied to other stars, the boundary conditions as mapped via spectropolarimetric observations and Zeeman splitting techniques \citep{Donati.Brown:97,Piskunov.Kochukhov:02}, are relatively unresolved. By determining the extent to which the Sun's Alfv\'en critical point is structured and non-spherical, stellar wind models with more accurate ram and magnetic pressure structure variation can be constructed. 
Improvement to this type of modeling has implications for stellar energetic particles transport \citep{Fraschetti.etal:22} and for galactic cosmic rays modulation and their penetration to inner astrospheres \citep{Herbst.etal:20}. 

Another perspective 
relates to exoplanet habitability. In highly magnetized stars (average surface magnetic field $\sim 0.5 -1$\,kG), the Alfv\'en surface might extend much farther out than the Sun, out to several tens of au 
\citep[e.g.,][]{Alvarado.Gomez.etal:22}. In addition, in several compact systems the planetary orbits are squeezed within $0.1$\,au 
\citep[e.g., TRAPPIST-1,][]{Gillon.etal:16}; thus, most of the planet orbits around the host star lie in sub-Alfv\'enic region (i.e., within the Alfv\'en surface), with dire consequences on the habitability \citep{Atkinson2024} and on the structure of the magnetosphere/ionosphere, especially for planets with no magnetic shielding. An unstructured Alfv\'en surface such as the one revealed by these Parker measurements, and the frequent transition from sub- to super-Alfv\'enic quiescent wind for close-in planets (smoothened by its thickness) requires sophisticated models of the atmospheric response to the wind ionization \citep{Gronoff.etal:20}. 

\section{Conclusions and Future Work}\label{sec:conclusions}

We draw the following conclusions:

\begin{itemize}
    \item Accounting for solar wind acceleration and mass--flux conservation is vital to accurately estimate the Alfv\'en surface height via scaling methods. 
    \item Estimating the height with multiple spacecraft in the inner heliosphere shows where the scaling methods are robust and where they are likely impacted by transients or time evolution.
    \item Sub-Alfv\'enic wind measured earlier by the Parker mission was related primarily to small outward extrusions in the Alfv\'en surface crossed longitudinally, while in the most recent encounters they are clearly sampled through radial evolution. Care should be taken to separate these physical circumstances when examining sub- vs. super-Alfv\'enic statistics since the outward extrusions may often be associated with transients such as CMEs.
    \item The solar Alfv\'en surface is farther from the Sun, less spherically symmetric and is rougher at solar maximum as compared to solar minimum. 
    \item In the current modestly strong Solar\,Cycle, the median height increases by approximately $30\%$ implying a near doubling of its angular momentum loss per unit mass--flux. 
    \item Accounting for the secular Solar\,Cycle variation in the Sun and other star's Alfv\'en surface height is vital for placing the Sun in stellar context, assessing angular momentum loss and spin down evolution. 
\end{itemize}

Moving forward, future perihelia from Parker Solar Probe at 9.86\,$R_\odot$ will be vital to collecting substantive statistics of sub-Alfv\'enic wind necessary to investigate any progression in physics above and below it. Based on these results, this will continue to be possible even as the Sun returns to solar minimum and its average Alfv\'en surface height retracts again. 

Our scaling method is quite general and will be able to provide a reasonable estimate of Alfv\'en surface shapes wherever measurements of solar wind speed and Alfv\'en speed are simultaneously available at or within 1\,au. It will therefore be of interest to extend the application of this work to probe the geometry of the Sun's Alfv\'en surface historically using older L1 data and even using Helios\,1 and 2 data down to 0.3\,au. Further, while in this work, the impact of latitude is generally neglected, it can be preserved in the mapping. This latter aspect will be of particular interest in extrapolating Solar Orbiter measurements inwards as it reaches progressively higher orbital inclination \citep[up to 30$^\circ$][]{Muller2020} to take this analysis from assuming co-planarity to constraining the Alfv\'en surface in 3D. This is currently only possible with solar wind modeling \citep{Chhiber2022}, historically with Ulysses data \citep{Verscharen2021} from much farther out than 1\,au, and is an outstanding goal of the recently launched Polarimeter to Unify the Corona and Heliosphere \citep[PUNCH;][]{Deforest2022}.

\vspace{5mm}
\textit{Acknoledgements} Parker Solar Probe was designed, built, and is now operated by the Johns Hopkins Applied Physics Laboratory as part of NASA’s Living with a Star (LWS) program (contract NNN06AA01C).  We thank the Solar Wind Electrons, Alphas, and Protons (SWEAP); team for providing data (PI: Michael Stevens, SAO) and FIELDS team for providing data (PI: Stuart D. Bale, UC Berkeley).The SWEAP Investigation and this publication are supported by the Parker Solar Probe mission under NASA contract NNN06AA01C. Solar Orbiter is a mission of international cooperation between ESA and NASA, operated by ESA. Solar Orbiter SWA data were derived from scientific sensors that were designed and created and are operated under funding provided by numerous contracts from UKSA, STFC, the Italian Space Agency, CNES, the French National Centre for Scientific Research, the Czech contribution to the ESA PRODEX programme and NASA. Solar Orbiter SWA work at the UCL/Mullard Space Science Laboratory is currently funded by UKSA/STFC grants UKRI919 and UKRI1204. Solar Orbiter magnetometer operations are funded by the UK Space Agency (grant UKRI943). We thank the ACE SWEPAM and MAG instrument teams and the ACE Science Center for providing the ACE data. We acknowledge the use of data from the NOAA Space Weather Prediction Center (2016): Deep Space Climate Observatory (DSCOVR). NOAA National Centers for Environmental Information. Dataset \url{http://doi.org/10.7289/V51Z42F7}. We also acknowledge the use of Wind spacecraft and the NASA/GSFC's Space Physics Data Facility's OMNIWeb (or CDAWeb or ftp) service. The CME catalog is generated and maintained at the CDAW Data Center by NASA and The Catholic University of America in cooperation with the Naval Research Laboratory. SOHO is a project of international cooperation between ESA and NASA. S.T.B, M.L.S, Y.J.R. T.N., K.W.P. and F.F. were partially supported by Parker Solar Probe project through the SAO/SWEAP subcontract 975569. K.G.K was supported in part by NASA grant 80NSSC24K0171. T.S.H. is supported by STFC grant ST/W001071/1. R.C. was supported in part by NASA/SWRI PUNCH subcontract N99054DS at the University of
Delaware. The authors acknowledge Michael McManus for writing up the first steps of the derivation in Appendix A.2 and A. J. Finley for discussion of their recent related work.
\vspace{5mm}

\software{This work made use of the version 7.0.0 of the Astropy (\url{http://www.astropy.org}) a community-developed core Python package and an ecosystem of tools and resources for astronomy  \citep{2013A&A...558A..33A, 2018AJ....156..123A, 2022ApJ...935..167A} and version 6.0.4 of the SunPy \citep{sunpy_community2020} open source software package \citep{stuart_j_mumford_2025_14919826}. We also acknowledge the use of the Python packages: Numpy \citep{harris2020array} version 1.26.4, Scipy \citep{2020SciPy-NMeth} version 1.13.1 and Matplotlib \citep{Hunter:2007} version 3.10.6. Initial data access was done using PySPEDAS version 1.7.1 \citep{Angelopoulos2019,Grimes2024}.
}
\vspace{8mm}

\begin{figure*}[ht!]
    \centering
    \includegraphics[width=0.9\linewidth]{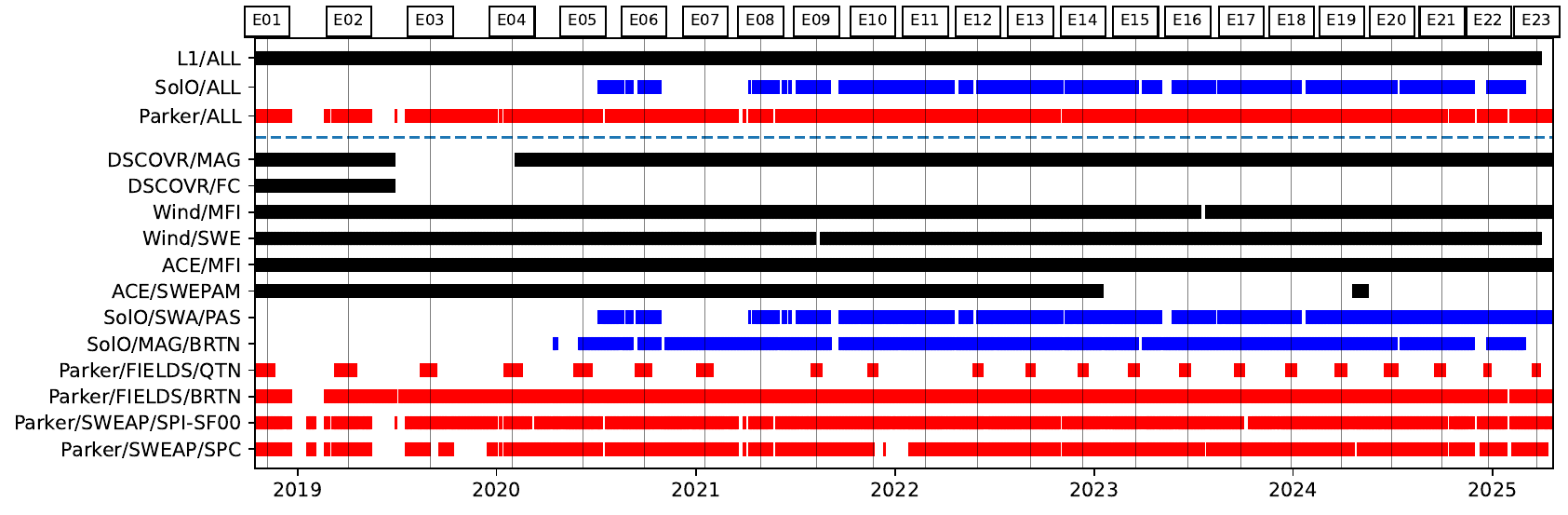}
    \caption{\textbf{Dataset coverage for producing $V_{SW}$ and $V_A$ estimates} For each instrument whose data is used in this work, a colored horizontal bar is shown with gaps where no data is available. The color scheme differentiates Parker, Solar Orbiter and L1-based measurements as in Figures\,\ref{fig:alfven-surface-topdown}--\ref{fig:roughness}. For each spacecraft/location, a summation bar at the top indicates overall coverage for both $V_{SW}$ and $V_A$. An inset panel indicates the Solar\,Cycle comparison and vertical bars and labels indicate the timestamps of each Parker Perihelion. 
}
    \label{fig:appendix:data-coverage}
\end{figure*}

\appendix
\section{Dataset and Preparation}\label{sec:appendix:data}
\subsection{Dataset Coverage}\label{subsec:appendix:coverage}

We estimate the 15-minute average of the radial component of the velocity ($V_{SW}$) and magnetic field ($B_R$), proton density ($N_p$) and proton temperature ($T_p$) measured by the following spacecraft: Parker Solar Probe \citep[Parker,][]{Fox2016}, Solar Orbiter \citep[SolO,][]{Muller2020}, Advance Composition Explorer ACE \citep[ACE,][]{Stone1998}, Deep Space Climate Observatory \citep[DSCOVR,][]{Burt2012}, and Wind \citep{1997AdSpR..20..559O, Wilson2021}. The dataset covers the time range starting on October\,2018 up to April\,2025.

For Parker, the plasma conditions ($V_{SW}$, $N_p$ and $T_p$) and magnetic field observations are obtained by the Solar Wind Electrons Alphas and Protons \citep[SWEAP,][]{Kasper2016} and Electromagnetic Fields Investigation Fields\,\citep[FIELDS,][]{Bale2016} instruments, respectively. The SWEAP suite comprises of a set of electrostatic analyzers: the {\it Solar Probe ANalyzers} for ions \citep[SPAN-i,][]{Livi2022} and for electrons \citep[SPAN-e][]{Whittlesey2020}, as well as a Faraday Cup the {\it Solar Probe Cup} \citep[SPC, ][]{Case2020}. The Radio Frequency Spectrometer \citep[RFS;][]{Pulupa2017} from FIELDS, also provides an independent measurement of the solar wind electron density \citep{2020ApJS..246...44M}.

These same observational quantities are obtained by Solar Orbiter with the Solar Wind Analyser \citep[SWA,][specifically the Proton-Alpha Sensor, PAS]{Owen2020} and magnetometer \citep[MAG,][]{Horbury2020}; from ACE by the Electron, Proton, and Alpha Monitor \citep[SWEPAM,][]{Gold98} and the magnetometer \citep[MAG,][]{1998SSRv...86..613S}; from DSCOVR by the Plasma-Magnetometer \citep[PlasMag, see][]{Burt2012} suite (with a electron spectrometer, a MAG - magnetometer and a FC - Faraday Cup); and in Wind by the Solar Wind Experiment \citep[SWE,][]{1995SSRv...71...55O} and the Magnetic Field Investigation \citep[MFI,][]{1995SSRv...71..207L} instruments.

To arrive at a ``best estimate'' of $V_{SW}$ and $V_A$ at each measurement point at a 15--minute cadence, the key dataset used in this work, we utilize as many available measurements as possible of each of the required basic parameters ($B_R$, $V_{SW}$ and $N_p$) and also the proton temperature, which is used to generate the acceleration profiles (see Appendix\,\ref{sec:appendix:solar-wind-modeling} below). 

The ingested data sources and availability over time used to provide individual measurements of $V_{SW}$ and $V_A$ are summarized in Figure\,\ref{fig:appendix:data-coverage} with data sources from Parker in {\it red}, for Solar Orbiter in {\it blue} and from different spacecraft at L1 in {\it black}. Solar Orbiter data only starts after the launch of the mission in early 2020. In the subsequent subsections, we describe these individual measurements and how they are combined. 

Before diving into details, it is worth briefly justifying our choice of 15 minute intervals. Our goal in this work is to examine large scale structure in the solar wind so that radial scaling methods can be applied. 15 minutes is chosen to be sufficiently long that no matter what distance from the Sun the measurement is taken, the sample is long enough to comfortably lie beyond the outer scale of MHD turbulence, avoiding any systematic changes in turbulence regimes between measurements.

Additionally, by taking measurement medians in these windows, we also avoid our results being distorted by Alfv\'enic fluctuations of at last magnetic and velocity fluctuations which tend to produce skewed distributions with long tails. The median of these distributions recovers the ``background'' plasma properties which are expected to smoothly evolve with distance from the Sun.

Lastly, 15 minutes is also sufficiently long that for all measurements discussed below, data products are available enabling, at minimum, hundreds of samples in each sample.

\subsubsection{Determination of $B_R$} \label{appendix:subsubsec:Br}

To determine $B_R$, we follow the ``Parker Spiral Method" \citep{Erdos2012,Erdos2014,Badman2021} in which we work with timeseries of the magnetic field vector as expressed in spherical coordinates ($|B|,\theta_B,\phi_B$). In these coordinates, the magnetic field components are approximately normally distributed (except for the azimuthal/Parker spiral angle which can still be skewed or bimodal, discussed further below) and therefore have an easily interpretable mean. On the other hand, raw measurements of the cartesian component, $B_R$, have skewed means which are highly dependent on the Parker spiral angle \citep{Badman2021}. Mean values of $B_R$ over 15 minute intervals are therefore constructed instead as: 
$$
<B_R> = <|B|> \sin\big(<\theta_B>\big)\cos\big(\phi_{B,P}\big),
$$

where $\phi_{B,P} = \tan^{-1} \big(\frac{\Omega_\odot R}{v_{SW}} \big)$ is taken as the mean theoretical Parker spiral angle in each interval and $<\theta_B>$ is the mean measured out-of-plane field orientation which is generally close to in-plane. By taking the Parker spiral angle from its theoretical dependence on $V_{SW}$ and $R$ instead of directly from measurements, we effectively remove instances where large field rotations mix together opposing polarity measurements resulting in artificially low apparent $B_R$ values which are not useful for radial scaling \citep[see Figures 3 \& 4 of][]{Badman2021}.

For Parker, measurements of the magnetic field come from the FIELDS instrument \citep{Bale2016} and the MAG-RTN-4 Sa/Cyc (~4.6\,Hz)  data product is used to produce large statistics in each 15--minute window. 

For Solar Orbiter, measurements come from the MAG instrument \citep{Horbury2020} via the 'MAG-RTN-NORMAL' product with a typical sampling rate of 8\,s.

Lastly, for L1, we make use of multiple spacecraft measuring the same quantity. Specifically, we utilize the Wind/MFI, ACE/MFI and DSCOVR/MAG instruments. We compute $<B_R>$ in 15 minute intervals as described above for each spacecraft individually, we then take the mean across all 3\,sources, approximating that they constitute measurements made at the same location (the exact Earth--Sun L1 point) to find a ``wisdom of the crowds'' estimate for this location in space. 

\subsubsection{Determination of $N_p$}

Parker Solar Probe plasma moment measurements are in general non-trivial to interpret due to an extremely variable aberration flow of the solar wind into the plasma detectors, as well as no individual instrument having complete field of view coverage due to spacecraft engineering considerations. 

We take advantage of the multiple independent measurements taken directly by the SWEAP instrument \citep{Kasper2016} via both the SPC \citep{Case2020} and SPAN-i \citep{Livi2022} sensors, and assuming quasi--neutrality, the electron density via quasi--thermal noise (QTN) measurements from FIELDS/RFS \citep{Bale2016, Pulupa2017, 2020ApJS..246...44M}. For SPAN-i, we first filter the data according to the 'EFLUX\_VS\_PHI' CDF variable to only select for density moments when the peak of the velocity distribution function is at least two instrument anodes from the edge of the detector \citep[a simple filtering method which collapses the VDF into one dimension, more sophisticated 2D methods are possible][]{Romeo_Thesis}.  For SPC, we filter according to the data quality flags (3,5,11,12,13,20,21,22,23) and select the full scan mode only \citep[see ][]{Case2020}. The end result is that for each 15--minute interval, we have a well formed distribution of measurements of $N_p$ from up to 3\,sources with poor quality data mostly excluded. For each of these distributions, we compute the median.

From these 3\,medians, we derive a best estimate of $N_p$ at Parker by following a simple algorithm of using the QTN measurement wherever available, and then the larger of the SPAN-i and SPC measurements. QTN data does not work when the plasma density is low, SPAN-i generally loses the peak of the VDF when the solar wind aberration flow is small, and SPC generally turns off close to the Sun due to instrument thermal issues. These factors conspire to produce a ``best'' source of the measurement broadly organized by heliocentric distance with QTN used at closest approach, SPAN-i used at intermediate distances and SPC used mostly outside of encounter mode. For the regions where multiple independent measurements are available, most (90\%) of intervals have a dispersion between measurements of less than 5\%, while the rest rarely exceed a dispersion of 10\%. 

For the proton density at Solar Orbiter, we use measurements by SWA/PAS \citep{Owen2020}. Although there is only one independent measurement available, SWA/PAS generally has a more complete view of velocity space than the ion sensors of Parker Solar Probe so is more straightforward to interpret. 

Finally, as with the magnetic field, the proton density at L1 is produced by independently estimating a median value in each 15--minute interval by instruments from multiple spacecraft, that is Wind/SWE, ACE/SWEPAM and (up until mid-2019) the DSCOVR Faraday Cup. 

\subsubsection{Determination of $V_{SW}$} \label{appendix:subsubsec:vsw}

The values of the radial solar wind speed in each 15--minute interval is determined from the same set of ion instruments as used for $N_p$ (with the exception of no equivalent measurement on Parker for QTN density). We apply the same filtering as described in relation to the density to build distributions of measurements from both instruments in 15--minute intervals and take the average of the median of both distributions. The filtering and SPC operational distances again mean that this measurement is largely powered by SPAN-i during perihelia, SPC outside of encounter, and with a small joint window during the inbound and outbound phases of the orbit. 

For Solar Orbiter and L1, the determination of $V_{SW}$ is identical to that of $N_p$. 

\subsubsection{Determination of $T_p$} \label{appendix:subsubsec:Np}

Proton temperatures are also estimated for this work although are only used indirectly through the isopoly acceleration profiles. In terms of their computation from the different spacecraft, for each case the approach is identical to that of $V_p$ with the one exception that for the SPAN-i measurements of $T_p$ we implement a method to reject the component of the temperature tensor which is impeded by the instrument's finite field of view (FOV). This procedure is fully described in the next Appendix section. 

A robust, statistically representative dataset of electron temperatures are not currently available available in the public data products of Parker Solar Probe and Solar Orbiter, so for subsequent sizing of electron thermal pressure gradients (discussed further in section \ref{sec:appendix:solar-wind-modeling}) we utilize prior work \citep{Dakeyo2022,Rivera2025}.

\subsubsection{Computation of $V_A$}

We close this appendix with a couple of notes about our computation of the Alfv\'en speed in this work. As reported in the main text, we compute this as:
\begin{align}
    V_A = B_R/\sqrt{\mu_0m_pN_p}
\end{align}

We note that this means we are stricly computing the ``radial'' Alfv\'en speed, that is, the component of the velocity of Alfv\'en waves (which are in general field aligned) in the radial direction. As the Parker spiral increases in inclination further from the Sun, this becomes more significantly different from the true field aligned Alfv\'en speed. We use this because the conservation of magnetic flux applies to the radial component of the magnetic field, and this simplifies the radial scaling behavior. 

The second note to point out is that we are computing only the Alfv\'en speed for protons in the solar wind, as evidenced by only including an $m_p N_p$ term. A more general computation would include contributions from alpha particles and heavier ions (and electrons in principle although their vanishingly small mass by comparison makes this a trivial correction):
\begin{align}
    V_A = B_R/\sqrt{\mu_0\sum_im_iN_i}
\end{align}

where $i$ denotes the different species present in the plasma.

Neglecting alpha particles does merit some discussion. Typical alpha abundances in the steady solar wind range from $1-5\%$, positively correlated with wind speed \citep[e.g.][]{Alterman2019}, and in transients such as during CMEs can typically reach up to 10\% and in extreme outlier cases up to 20\% \citep{Johnson2024}. Because the mass density of alphas is quadruple that of the same number of protons, a relatively small abundance of alpha particles can still have a non-negligible impact on the computation of the Alfv\'en speed.

Taking the fast wind 5\% value as a typical worst case value for the steady streams most important to this study, we se in this case that the correction to the mass density would be an increase of 20\%. Including this in the computation of the Alfv\'en speed yields a ~9\% potential reduction to the Alfv\'en speed. In the analysis presented in this work, this translates to our estimates of the surface of the Alfv\'en surface being a lower bound with a correction of order $\sim$1$R_\odot$ for fast wind and even smaller for slow wind, i.e. comparable but somewhat smaller than the  general range of variability observed in this work of 3-8 $R_\odot$ (Figure\,\ref{fig:roughness}. A future study which incorporates the alpha abundance would be interesting to investigate if any secular changes in stream types and therefore alpha abundances have a further systematic correction on the Alfv\'en surface studied here.  

\subsection{Parker/SPAN-i Proton Temperature Correction} \label{appendix:subsec:Tp}

\begin{figure}[h!]
    \centering
    \includegraphics[width=0.8\linewidth]{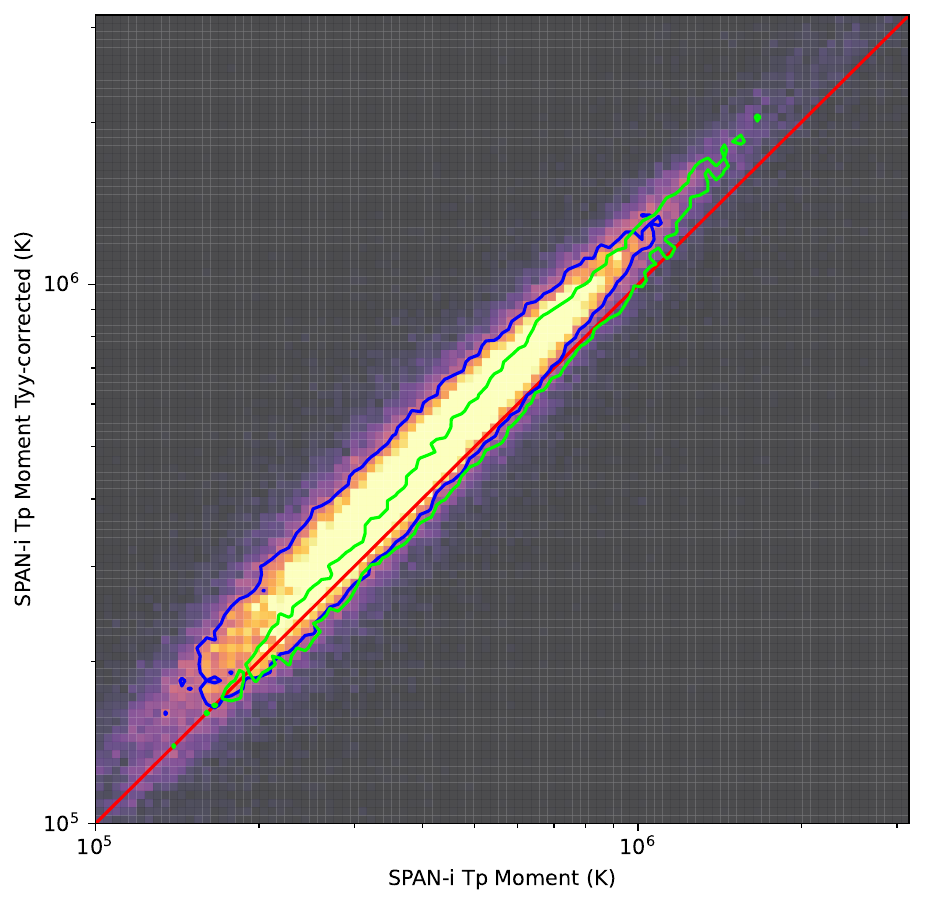}
    \caption{\textbf{Effect of SPAN-i Temperature Correction}. A 2D heatmap of the “gyro-corrected” scalar proton temperature vs the original SPAN-i moment 15--minute medians. A {\it red} diagonal line indicates y=x. A {\it blue} contour shows the corrected temperature is systematically hotter in general. A {\it green} contour shows the distribution and correction is less severe after selecting for good field of view intervals, however at the highest temperatures the systematic increase is still observed. }
    \label{fig:Tp-Correction}
\end{figure}

In addition to proton density and velocity, this work also utilizes statistical trends in the solar wind proton temperature to provide thermal pressure gradients for the isopoly models discussed in the next Appendix section. For the closest approaches of Parker Solar Probe, which are central to deriving these constraints, we use data from SWEAP/SPAN-i \citep{Livi2022}. SPAN-i L3 data reports a scalar temperature which is the trace of the temperature tensor  computed in instrument coordinates divided by 3. Some of these tensor components (specifically those involving the instrument-frame Y-coordinate) are systematically affected by instrument finite field of view affects. In this Appendix section, we show that by making a simple gyrotropic assumption for the form of the temperature tensor, we can ignore these tensor components and recompute the scalar temperature without this distortion.   \\ 


We first define the rotation matrix $\boldsymbol{R}_B$ via the expression: 

\begin{equation*}
\boldsymbol{R}_B B \hat{e}_{z} = \begin{pmatrix}
a&b&c\\
d&e&f\\
g&h&i\\
\end{pmatrix}B \hat{e}_{z} =  \boldsymbol{B}_{INST},
\end{equation*}

which acts to rotate any vector aligned with the SPAN-i instrument frame z axis to be aligned with the magnetic field vector in the instrument frame, $\boldsymbol{B}_{INST}$. We next assume a gyrotropic temperature tensor such that in a coordinate frame with the z-axis aligned with the magnetic field, we have:

\begin{equation*}
\boldsymbol{T}_{B} = \begin{pmatrix}
T_\perp & 0 & 0 \\
0 & T_\perp & 0 \\
0 & 0 & T_\parallel \\
\end{pmatrix},
\end{equation*}
and a measured temperature tensor in the instrument frame: 
\begin{equation*}
\boldsymbol{T}_{INST} = \begin{pmatrix}
T_{xx} & T_{xy} & T_{xz} \\
T_{xy} & T_{yy} & T_{yz} \\
T_{xz} & T_{yz} & T_{zz} \\
\end{pmatrix},
\end{equation*}
where $\boldsymbol{T}_{INST}=\boldsymbol{T}_{INST}^T$ is a symmetric matrix. These latter two matrices are related by the coordinate transformation: 
$$ \boldsymbol{T}_{B} = \boldsymbol{R}_B \cdot \boldsymbol{T}_{INST} \cdot \boldsymbol{R}_B^T. $$

We now can write down equations for all tensor components which do not depend on the SPAN-i y coordinate:
$$
T_{xx} = (a^2 + d^2)T_\perp + g^2 T_\parallel, \\
$$
$$
T_{xz} = (ac + df)T_\perp + gi T_\parallel, \\
$$
$$
T_{zz} = (c^2 + f^2)T_\perp + i^2 T_\parallel. \\
$$

Using the fact that the rows and columns of rotation matrices are mutually orthogonal unit vectors, these equations can be further simplified to yield:

\begin{align*}
T_{xx} &= T_\perp + g^2 (T_\parallel - T_\perp), \\
T_{xz} &= gi (T_\parallel - T_\perp), \\
T_{zz} &= T_\perp + i^2 (T_\parallel - T_\perp), \\
\end{align*}

which is separable and solvable to get expressions for $T_\perp$ and $T_\parallel$:
\begin{equation*}
T_{xx} = T_\perp + (g/i) T_{xz} \implies T_\perp =  T_{xx} - (g/i) T_{xz}, \\ 
T_{zz} = T_\perp + (i/g) T_{xz}  \implies T_\perp =  T_{zz} - (i/g) T_{xz}, \\
\end{equation*}
and therefore, 
\begin{equation*} 
T_\parallel = T_\perp + \frac{T_{xz}}{gi} = T_{xx} + \frac{g}{i}\bigg(1 - \frac{1}{g^2}\bigg) T_{xz}. 
\end{equation*}

We note that since there are two equations for $T_\perp$ this implies an additional constraint which is essentially a measure of how good our gyrotropic assumption is: 
\begin{align}
T_{xx} - T_{zz} = \bigg(\frac{g}{i} - \frac{i}{g}\bigg) T_{xz}.
\end{align}\label{Eqn:constraint}

Since the rotation matrix components are determined by the magnetic field vector in the instrument frame, we can obtain $g$ and $i$ in terms of the magnetic field vector measured in the instrument frame.

To do this, we write the magnetic field measurement as $\boldsymbol{B}_{INST} = (B_x, B_y, B_z)^T = B \hat{b}$. The rotation that aligns the instrument frame z-axis with this vector can be written in an axis-angle formulation. The relevant angle is defined as $\cos \theta = \hat{b} \cdot \hat{z} = B_z/B$, and the axis of rotation is $\hat{k} = \hat{z} \times \hat{b} = \frac{1}{B} (- B_y, B_x, 0)^T$. In this form, an arbitrary rotation matrix is given by:
\begin{equation*}
\boldsymbol{R} = \cos \theta \boldsymbol{I} + \sin \theta [\hat{k}]_\times + (1-\cos \theta) \hat{k} \hat{k}, 
\end{equation*}
which in our case can be written as,
\begin{align*}
\boldsymbol{R_B} = B_z/B \boldsymbol{I} + \sqrt{B_x^2 + B_y^2} (1/B^2)\begin{pmatrix}
0 & 0 & B_x\\
0 & 0 & B_y\\
-B_x & -B_y &0\\
\end{pmatrix} + \\ (1 - B_z/B)(1/B^2) 
\begin{pmatrix} 
B_y^2 & -B_y B_x & 0 \\
-B_y B_x & B_x^2 & 0 \\
0 & 0 & 0 \\
\end{pmatrix}.
\end{align*}

We only need to know the $g$ and $i$ elements which we can read off as: 

\begin{align*}
g &=  -\frac{B_x}{B} \sin \theta = (-B_x/B^2) \sqrt{B_x^2 + B_y^2}  \\
i &= \cos \theta = \frac{B_z}{B},
\end{align*}

which finally can be manipulated to obtain:
\begin{align}
\boxed{T_\perp = T_{zz} + \frac{T_{xx} - T_{zz}}{1-(B_x/B)^2\tan^2 \theta}} \label{eqn:Tpar} \\
\boxed{T_\parallel = T_\perp + \frac{T_{xx} - T_{zz}}{(B_x/B)^2\sin^2 \theta - \cos^2 \theta}} \label{eqn:Tperp}
\end{align}

To illustrate the effect of this correction on our measurements, in Figure\,\ref{fig:Tp-Correction} we present a comparison between the SPAN-i L3 scalar temperature on the x-axis, and this gyrotropic correction on the y-axis. A heatmap and {\it blue} contour shows the correction applied to all data, while the {\it green} contour shows how the distribution shifts when a field of view criterion is applied via requiring the peak of the velocity distribution function (VDF) be at least two instrument anodes into the field of view, which is a useful way of rejecting many VDFs which are impeded by the Parker heat shield. We see in both cases, this correction yields a slightly higher scalar temperature and a larger correction at higher temperatures. Pre-filtering with a FOV criterion reduces the needed correction significantly, but still does not exactly remove it at high temperatures. This plot sanity checks the correction as we expect the tensor component this method removes to be artificially lower than the true temperature component due to the VDF being truncated in that direction. Further, we expect the effect to be worsened for hotter temperatures when the wings of the VDF are more significantly impeded by the field of view. Lastly, the reduction in needed correction after filtering by FOV also makes sense as it increases the likelihood that, especially for cool temperatures, the whole VDF is collected by the SPAN-i detector.

Although not fully exploited here, we note that Equations \ref{eqn:Tpar} and \ref{eqn:Tperp} are an efficient and direct way to estimate a gyrotropic decomposition of the temperature tensor measured by SPAN-i, as compared to traditional reconstruction methods via fitting bi-maxwellians \citep[e.g.][]{Huang2020,Woodham2021} or more sophisticated decompositions such as using Slepian basis functions \citep{Das2025}. 

However, there are certain limitations. First, the form of the gyrotropic tensor is an assumption which amounts to presuming the VDF is a prolate ellipsoid oriented along the magnetic field. It does not allow for different temperatures in the two perpendicular directions, and does not test for any departure of the real VDF from this idealized assumption, although the additional constrain developed earlier in equation \ref{Eqn:constraint} can be used to assess this.

Second, the mathematical solutions in Equations \ref{eqn:Tpar} and \ref{eqn:Tperp} do diverge under certain conditions (when the denominators go to zero). This is primarily determined by the orientation of the magnetic field interacting with the component of the tensor which is being thrown away in this method. Specifically, $T_\parallel$ is undetermined if the magnetic field is aligned with the instrument y-axis.

\section{Parameterization of ``Isopoly'' Solar Wind Models}\label{sec:appendix:solar-wind-modeling}

\begin{figure}[h!]
    \centering
    \includegraphics[width=\linewidth]{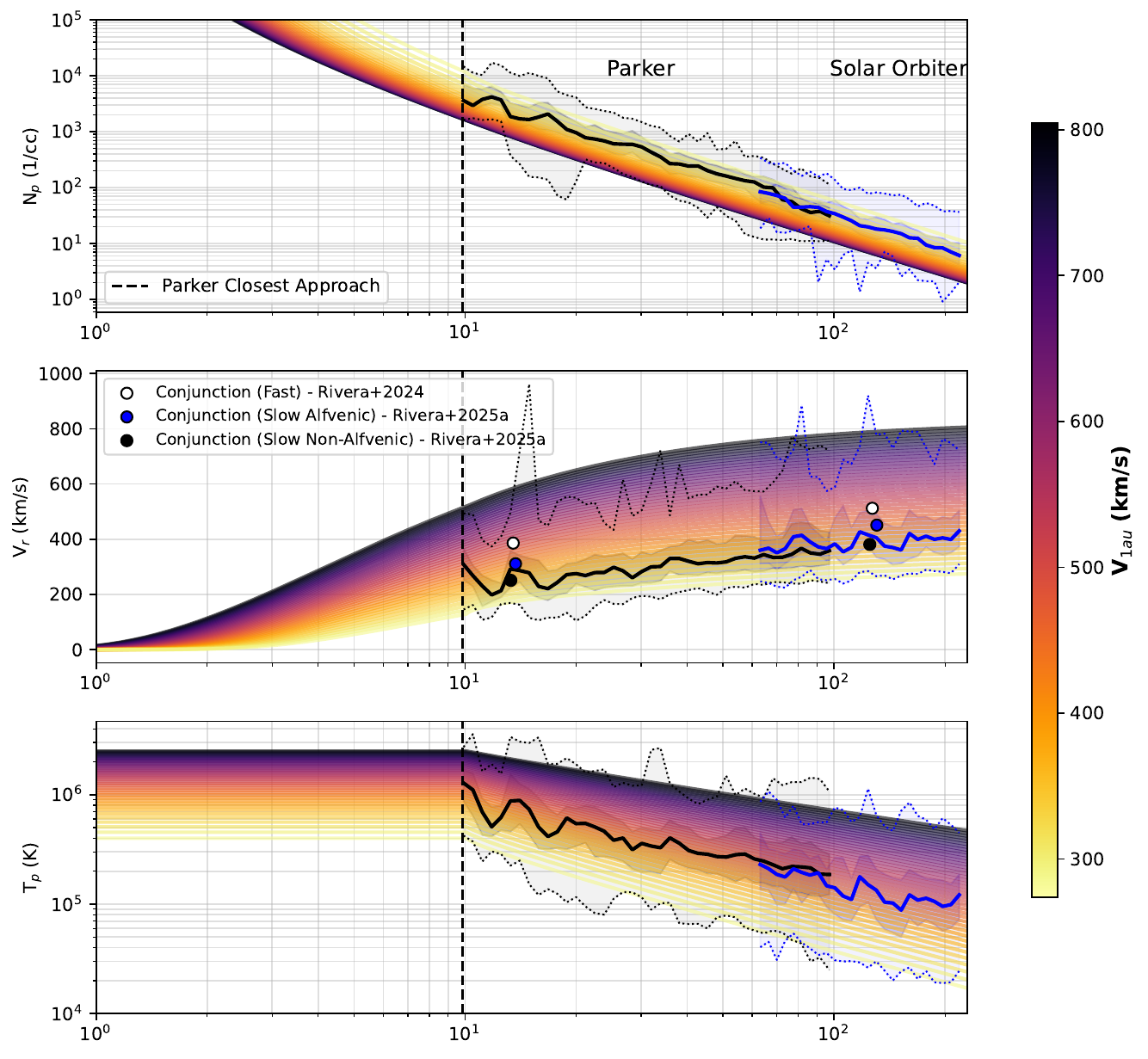}
    \caption{\textbf{``Isopoly’’ wind profiles and data statistics}. From {\it top} to {\it bottom}, the panels show isopoly profiles of proton density, proton velocity, and proton temperature, colorized as in Figure\,\ref{fig:intersection} according to the wind speed at 1\,au. These models are superimposed on statistics of the 15--minute cadence data set of each quantity as a function of radial distance. In each panel, the solid line shows the median as a function of distance, while two progressively fainter regions annotate the interquantile range and the 5/95 percentile ranges respectively. {\it Black} ({\it blue}) lines and shading indicate Parker and Solar Orbiter statistics respectively. In the {\it middle} panel, scatter points provide additional context depicting conjunction results on solar wind acceleration from \citep{Rivera2024a, Rivera2025}. A dashed vertical bar indicates Parker's closest perihelion distance.
}
    \label{fig:model-profiles}
\end{figure}

In this work, we make use of a set statistically justified acceleration profiles and resulting mass flux profiles to produce radial scaling of $V_{SW}$ and $V_A$. The profiles come from two-fluid (electron and proton) ``isopoly'' models \citep{Dakeyo2022} with an additional external force profile \citep[][]{Shi2022, Rivera2024a, Rivera2025} to achieve the fastest asymptotic speeds. 

These models produce Parker solar wind acceleration solutions \citep{Parker1958, Parker1960} given a prescribed isothermal coronal electron and proton temperature ($T_e$ and $T_p$ respectively), a height $R_{iso}$ where the temperature profiles depart from being approximately isothermal and instead cool with polytropic indices $\gamma_e$ and $\gamma_p$ respectively. 

These parameters are set via a combination of prior work and an empirical examination of the statistical behavior with respect to heliocentric distance of the 15--minute in situ dataset used in this work, discussed further below. 

This statistical behavior and the resulting isopoly profiles are illustrated in Figure\,\ref{fig:model-profiles} where proton density, velocity and temperature profiles are plotted colorized according to asymptotic wind speed at 1\,au as in Figure\,\ref{fig:intersection}. In each case, the statistical datasets of these same quantities from Parker and Solar Orbiter are plotted in {\it gray} and {\it blue} curves respectively. For both spacecraft, a {\it black} ({\it blue}) solid curve shows the median of each quantity vs distance, while progressively fainter shaded regions show the interquartile range in each distance bin as well as the 1st and 99th percentiles. In the middle panel which shows velocity, scatter points show the acceleration from Parker to Solar Orbiter analyzed in \citet{Rivera2024a,Rivera2025}, consistent with the middle range of our acceleration profiles. 

The median and percentiles of the Parker and Solar Orbiter data are largely contiguous and mutually consistent for the small region of heliocentric distance for which they overlap, demonstrating the statistics are quite well sampled and not skewed by the differing orbit and sampling time periods of the two missions. 

The isopoly curves span the 1st-99th percentiles of each data quantity over almost all heliocentric distances, and general correlations are preserved in both the models and statistical data. Specifically, fast wind has consistently lower densities, higher temperatures and a slower fall off with distance, while the slow wind is denser, cooler and cooling more quickly (closer to adiabatic expansion), and this is all consistent with \citet{Dakeyo2022}. We therefore argue that this set of isopoly profiles are a good representation of the acceleration and mass--flux profiles across most types of solar wind at least out to 1\,au and are therefore useful and usable for the analysis presented in the main text of this work. Moreover, the consistent Alfv\'en surface localization presented in Figures\,\ref{fig:methods:validation} and \ref{fig:alfven-surface-topdown} when scaled in from 1\,au and from nearer to the Sun, is further evidence that these profiles are a good representation of the radial evolution of the solar wind. 

We close with a brief summary of the parameter ranges used to produce the curves shown here, along with a brief rationale for setting these ranges.

\subsection{Isopoly Parameters}

The full set of parameters for a given wind profile are $\{R_{iso}, T_p, T_e,\gamma_p,\gamma_e, F(R)\}$ which are respectively the distance of the boundary between the isothermal and polytropic portion of these profiles, the proton and electron isothermal temperatures, the proton and electron polytropic indices and the external forcing profile. The family of curves comes from setting an upper and lower limit and linearly sampling 40\,values between these limits in each case. 

For each parameter, we report in Table\,\ref{table:appendix:parameter-val} these bookends corresponding to the slowest and fastest winds. $T_p$, $\gamma_p$ and $R_{iso}$ are set to produce the set of proton temperature curves shown in the bottom panel of Figure\,\ref{fig:model-profiles}. For simplicity, we set $R_{iso}$ to 10\,$R_\odot$, and then the ranges of $T_{p}$ and $\gamma_p$ are sized to the 1st and 99th percentiles of the data in the heliosphere. This results in consistent behavior with the statistical fits of \citet{Dakeyo2022} for which the faster (slower) wind has a shallower (steeper) polytropic index, higher (lower) isothermal temperature, and further yields a coronal proton temperature for fast wind of 2.5\,MK consistent with UVCS polar coronal hole observations \citep{Cranmer2020b}.

With well-vetted measured electron temperature profiles out of scope for the present work, the electron polytropic indices are set to those reported by \citet{Dakeyo2022}, while the coronal temperature, $T_e$ is set to 1\,MK in line with \citet{Rivera2024a} and \citet{Cranmer2020b}. We check that the resulting slowest wind speed profile matches the 1st percentile of the acceleration statistics (middle panel of Figure\,\ref{fig:model-profiles}) where thermal pressure gradients are expected to fully explain the acceleration \citep{Halekas2020,Halekas2023,Alterman2025}.

Lastly, the external force reuses the analytic function from \citet{Shi2022,Rivera2024a}:
\begin{align}
    F(R) = f_0 \frac{1+ \beta(R/R_\odot-1)}{(R/R_\odot)^2} e^{\alpha(1-R_\odot/R))},
\end{align}
where we vary the strength parameter $f_0$ to change the size of the force according to asymptotic wind speed. We set the other parameters fixed at $[\alpha=8,\beta=8]$. This produces a slightly stronger force profile at lower altitudes compared to the parameters used in \citet{Rivera2024a} $[\alpha=0.2,\beta=74]$ which better matches the acceleration for the fastest 99th percentile of statistical measurements while leaving the slower and intermediate speed profiles relatively unchanged. 

The parameter $f_0$ is then sized to span from 0 for the slowest speed winds to the maximum shown in Table\,\ref{table:appendix:parameter-val} such that the fastest profile match the 99th percentile of wind speeds at 10\,$R_\odot$ and 1\,au (215\,$R_\odot$), reaching around 800\,km\,s$^{-1}$. Further, the intermediate acceleration profiles are checked for consistency with the acceleration profiles of \citet{Rivera2024a, Rivera2025}.

The code to produce these profiles given the above parameters is available at \url{https://github.com/STBadman/ParkerSolarWind}. We note the coronal behavior here is poorly constrained and effects such as non-radial flux tube expansion \citep{Dakeyo2024b} and non-isothermal coronal temperatures \citep{Dakeyo2025} are likely important low in the corona. However, for the purpose of studying radial scaling around and outwards from the Alfv\'en surface which is almost always exterior to Parker's closest approaches (see Figure\,\ref{fig:alfven-surface-topdown}), the profiles are well constrained by in situ data. 

\begin{table*}
\begin{tabular}{|c|c|c|}
\hline
\textbf{Parameter Name} & \textbf{Slowest Wind} ($\sim$250\,km\,s$^{-1}$ @ 1\,au) & \textbf{Fastest Wind}  ($\sim$810\,km\,s$^{-1}$ @ 1\,au)  \\ \hline
$T_p$ (MK) & 0.4 & 2.5  \\ \hline
$T_e$ (MK) & 1.0 & 1.0  \\ \hline
$\gamma_p$  & 1.45 & 1.25 \\ \hline
$\gamma_e$ & 1.2 & 1.3 \\ \hline
$R_{iso}$ ($R_\odot$) & 10 & 10 \\ \hline
$f_0$ ($GM_\odot/R^2_\odot$) & 0 & 3.3$\times 10^{-4}$ \\ \hline
\end{tabular}\label{table:appendix:parameter-val}
\caption{Isopoly Parameter Ranges}
\end{table*}

\bibliographystyle{aasjournal}

\end{document}